\documentclass{ws-ijmpe}

\begin{document}

\markboth{Authors' Names}{Instructions for Typing Manuscripts
(Paper's Title)}

\catchline{}{}{}{}{}

\title{ON THE TFD TREATMENT OF COLLECTIVE VIBRATIONS IN HOT NUCLEI}

\author{\footnotesize ALAN A. DZHIOEV}

\address{Bogoliubov Laboratory of Theoretical Physics,\\ Joint
Institute for Nuclear Research, Dubna 141980, Russia\\
dzhioev@theor.jinr.ru}

\author{A. I. VDOVIN}

\address{Bogoliubov Laboratory of Theoretical Physics,\\ Joint
Institute for Nuclear Research, Dubna 141980, Russia\\
vdovin@theor.jinr.ru}

\maketitle

\begin{history}
\received{(received date)} \revised{(revised date)}
\end{history}

\begin{abstract}
The approach in a theory of collective excitations in hot nuclei
exploring the formalism of thermo field dynamics and the model
Hamiltonian consisting of a mean field, the BCS paring interaction
and long-range particle-hole effective forces is reexamined. In
contrast with earlier studies it is found that a wave function of
a thermal phonon is depended not only on the Fermi-Dirac thermal
occupation numbers of Bogoliubov quasiparticles consisting the
phonon but on the Bose thermal occupation numbers of the phonon as
well. This strongly affects a thermal phonon couplings due to
renormalizing of a phonon-phonon interaction and  enlarging the
number of thermal two-phonon configurations coupled with
one-phonon ones. Moreover, it is shown that the formulation of the
double tilde conjugation rule for fermions proposed by I.~Ojima is
more appropriate in the context of the present study than the
original one by H.~Umezawa and coworkers.
\end{abstract}

\section{Introduction}

The present paper addresses some aspects of theoretical treatment
of collective excitations of a hot nucleus within the thermo field
dynamics (TFD).

The TFD\cite{Umezawa75,Umezawa82} is known as a powerful tool in studying many-body problems at
finite temperatures. There exist numerous TFD applications in condensed matter and high energy
physics (see e.g.\cite{Umezawa82,Arimitsu91,Henning95,Arimitsu99}). In spite of several attractive
properties the TFD formalism seems to be less popular in the nuclear theorist community. During the
last 20 years only a dozen works dealing with the TFD application to nuclear structure problems
were
published.\cite{Tana88,Hats89,Tana90,Civi93,Tana94,Kos94,Kos95,Kos96,Avde96,Civi98,Storo99,Storo01,Storo04,Tana05}

An important step in application of the TFD formalism for extending the standard nuclear theory
methods (the HFB approximation, RPA, a boson expansion technique) to finite temperatures was made
by T.~Hatsuda.\cite{Hats89} In,\cite{Tana88,Tana94,Avde96,Storo99,Storo01} different approximations
going beyond the thermal RPA were formulated. O.~Civitarese and coauthors\cite{Civi93,Civi98}
applied the TFD to problems of treating nuclear pairing in the BCS and the random phase
approximations as well as quasi-continuum (or resonant) single-particle states at finite
temperatures. Moreover, the TFD was used to prove the Bloch-Messiah theorem at finite
temperatures\cite{Tana90} and construct number-projection methods for the BCS pairing in hot
nuclei.\cite{Kos96,Tana05}

The methods and approximations developed in the papers cited above
were not applied to numerical calculations of any nuclear
properties. Only simple solvable models were used to demonstrate
the new achievements. At the same time, in\cite{Kos94,Kos95} the
TFD was combined with the Quasiparticle-Phonon nuclear model
(QPM)\cite{Soloviev92} to formulate an approach allowing one to
treat microscopically a coupling of different excitation modes in
hot nuclei. This coupling is known to be responsible for the
spreading width of giant resonances in cold spherical nuclei, and
the TFD-QPM approach was used to analyze a thermal behavior of a
giant dipole resonance width.\cite{Storo04}

It appeared that in contrast with earlier studies of the problem\cite{Bort86} (see
also\cite{Seva97}) based on the Matsubara Green function technique and the Nuclear Field
Theory,\cite{NFT77} the quasiparticle-phonon interaction at finite temperature in the TFD-QPM
approach did not depend on the thermal occupation numbers of phonons. Only the Fermi-Dirac thermal
occupation numbers of noninteracting BCS quasiparticles appeared in corresponding formulae of
Refs.\cite{Kos94,Storo04} Some aspects of this difference were discussed in\cite{Storo04} (see
also\cite{Aouissat01}).

Recently, studying weak transitions of the Gamow-Teller type in
hot nuclei\cite{Dzhioev08} within the TFD  we found some
inconsistencies in the TFD-QPM approach of\cite{Kos94,Kos95}
related to structures of thermal vibrational phonons. Moreover,
new points in the TFD-QPM approach appear if one adopts changes in
the TFD formulation proposed by Izumi Ojima\cite{Ojima81,Mats85}
that concern the alternative formulation of the double tilde
conjugation rule.

The reasons given above compel us to reexamine the TFD-QPM
approach of\cite{Kos94,Kos95,Storo04} to a quasiparticle-phonon
coupling at finite temperatures. This is the main aim of the
present paper.

The paper is  organized as follows. A brief summary of the TFD formalism is given in
Sec.~\ref{TFD_formalism}. This seems to be necessary because we use here a variant of the theory
which somewhat differs from the standard one.\cite{Umezawa75,Umezawa82,Hats89} In Sec.~\ref{main},
the TFD formalism is applied to the nuclear Hamiltonian consisting of a mean field, the BCS pairing
interaction and separable particle-hole forces, i.e., to the QPM Hamiltonian.\cite{Soloviev92} In
Subsec.~\ref{qp_approx}, the thermal BCS pairing is considered. This part of the paper is close to
that of,\cite{Civi93,Kos94,Kos95} a discussion of $\gamma$-transitions between a thermal vacuum
state and thermal quasiparticle excitations is the only addition. In Subsec.~\ref{RPA_approx}, the
equations of the thermal random phase approximation (TRPA) are derived. It is shown why and how the
thermal Bose-Einstein occupation factors should appear in the expressions for thermal quasiparticle
amplitudes of a thermal phonon wave function. The points missed in\cite{Kos94,Kos95,Storo04} are
analyzed and discussed. The new formulae of the quasiparticle-phonon coupling term at finite
temperature are evaluated in Subsec.~\ref{high_approx}. The summary and conclusions are given in
Sec.\ref{concl}.

\section{The TFD formalism} \label{TFD_formalism}

We consider a system of Fermi particles at finite temperature and
treat its statistical properties within the grand canonical
ensemble. In the standard statistical mechanics a heated system is
described by a mixed state density matrix $\rho$ which is the
solution to the Liouville - von~Neumann equation
\begin{equation}\label{Liouville}
i\,\frac{\partial\rho}{\partial t}=[H,\rho].
\end{equation}
Here $H$ is the Hamiltonian of the system under consideration with
eigenstates $|n\rangle$ and eigenvalues $E_n$ (the chemical
potential is included in $H$). The thermal average of an arbitrary
operator $A$ is given by
\begin{equation}\label{exp_val_st}
   \langle\!\langle A\rangle\!\rangle={\rm Tr}[\rho A]/{\rm Tr}[\rho]=
\sum_n {\rm e}^{-E_n/T}\langle n|A|n\rangle \Bigl/ \sum_n {\rm
e}^{-E_n/T},
\end{equation}
where  $T$ is the temperature in units of energy. The main idea
behind TFD is to define a special state which is named a thermal
vacuum $|0(T)\rangle$, such that the thermal average of $A$ equals
the expectation value of $A$ with respect to this state
\begin{equation}\label{exp_val}
   \langle\!\langle A\rangle\!\rangle=\langle0(T)|A|\,0(T)\rangle~.
\end{equation}

To construct $|0(T)\rangle$, one should double the Hilbert space
of the system by adding the so-called tilde states
$|\widetilde{n}\rangle$.\cite{Umezawa75} These tilde-states are
the eigenstates of the tilde Hamiltonian $\widetilde H$ with the
same eigenvalues as $H$, i.e. $\widetilde
H|\widetilde{n}\rangle=E_n|\widetilde n\rangle$. Thus, the Hilbert
space of a heated system is twice as large as of the corresponding
initial (cold) one.

In the enlarged space covered by the states $|n\rangle \otimes |\widetilde m\rangle$ two types of
operators are acting -- the ordinary, say, $A_i$ and the tilde ones $\widetilde A_i$. There is a
one-to-one correspondence between these two sets $A_i \leftrightarrow \widetilde A_i$. However,
ordinary operators can change only ordinary states $|n\rangle$ whereas tilde operators change only
tilde-states $|\widetilde n\rangle$. There exist the following rules of tilde conjugation
operation:\cite{Umezawa75}
\begin{eqnarray}\label{tc_rule}
\widetilde{(A_1A_2)}&=&\widetilde{A}_1\widetilde{A}_2,\nonumber\\
\widetilde{(c_1A_1+c_2A_2)}&=&c_1^*\widetilde{A}_1+c_2^*\widetilde{A}_2,
\end{eqnarray}
where $c_1$, $c_2$ are $c$-numbers. The asterisk denotes the
complex conjugation.  The tilde operation is supposed to commute
with the Hermitian conjugation
\begin{equation}
(\widetilde{A})^\dag=\widetilde{A^\dag}~,
\end{equation}
Moreover, it is required that ordinary-
and tilde-operators should commute or anticommute with each other
\begin{equation}
[\widetilde{A_1},A_2]_{\mp}=0
\end{equation}
depending on their bosonic or fermionic nature.

Introduction of the doubled Hilbert space of a heated system
enables one to construct the above thermal vacuum $|0(T)\rangle$
in the following way:
\begin{equation}\label{T_vac}
    |0(T)\rangle=
    \sum_n {\rm e}^{-E_n/2T}{\rm e}^{i\alpha_n}
    |n \rangle \otimes |\widetilde n\rangle \Bigl/ \sqrt{\sum_n {\rm e}^{-E_n/T}},~~~(\alpha_n\in \mathbb{R}).
\end{equation}
Hereafter, the thermal vacuum (\ref{T_vac}) will be referred to as
the exact thermal vacuum. The thermal vacuum is invariant under
the tilde operation.

Till now our discussion follows the line of.\cite{Umezawa75,Umezawa82} The new point appears in the
definition of the so-called double tilde conjugation rule (DTCR). Originally,\cite{Umezawa75} DTCR
was introduced in the form
\begin{equation}
\widetilde{\widetilde A} = \rho_A\, A,
\end{equation}
where $\rho_A=1$ if $A$ is a bosonic operator and $\rho_A=-1$ if $A$ is a fermionic one. This DTCR
form was used in all of the TFD applications to nuclear structure problems, i.e., in.
\cite{Tana88,Hats89,Tana90,Civi93,Tana94,Kos94,Kos95,Kos96,Avde96,Civi98,Storo99,Storo01,Storo04,Tana05}
 However, quite long ago the other form of DTCR was proposed by I.~Ojima,\cite{Ojima81} namely,
\begin{equation}\label{Ojima_dtcr}
\widetilde{\widetilde A} = A.
\end{equation}
That is, the Ojima's version of DTCR does not distinguish between
bosonic and fermionic operators. This form is based on the
equivalence between the algebraic structure of TFD and that of the
axiomatic statistical mechanics ($c^*$-algebra approach)
established by I.~Ojima. In the present paper, we will compare
consequences of using the two DTCR versions. For that we define
DTCR for a fermionic $A$ in the following way:
\begin{equation}\label{DTC_rule}
\widetilde{\widetilde A}= -\sigma^2 A~,
\end{equation}
where  $\sigma$ is equal to either 1 ($\widetilde{\widetilde
A}=-A$)  or $i$ ($\widetilde{\widetilde A}=A$). Thus, the DTCR
definition (\ref{DTC_rule}) includes both the above DTCR variants.

The Schr\"{o}dinger equation for a hot system in the doubled
Hilbert space reads
\begin{equation}\label{Shr}
 i\frac {\partial}{\partial
t}|\Psi(t,T)\rangle={\cal H}|\Psi(t,T)\rangle,
\end{equation}
since  ${\cal H}=H -\widetilde H$ is the time-translation operator
and, therefore, the Hamiltonian of a hot system. The operator
$\mathcal H$ is named the thermal Hamiltonian.\cite{Umezawa75} The
exact thermal vacuum (\ref{T_vac}) is the eigenstate of $\mathcal
H$ corresponding to the zero eigenvalue. Equation~(\ref{Shr})
describes the time evolution of a system at finite temperature
and, thus, is the analog of the Liouville - von~Neumann
equation~(\ref{Liouville}).

The properties of elementary excitations of the system at $T\ne0$
are determined by ${\cal H}$. Excitation energies of various modes
at $T \neq 0$ are the eigenvalues of ${\cal H}$ but not of the
original Hamiltonian $H$ and in general they depend on $T$.
Moreover, any eigenstate of $\cal H$ with positive energy has its
counterpart -- the tilde-conjugate eigenstate with negative
energy. Following H.~Umezawa  we consider creation of a
tilde-state with negative energy as annihilation of a thermally
excited state. This is a way to treat  excitation and
de-excitation processes in a heated quantum system within TFD.

Whereas the dynamical development of the system is carried by the
thermal Hamiltonian, its thermal behaviour is controlled by the
thermal vacuum. The grand canonical value of an observable
corresponding to operator $A$ should be calculated as $\langle
0(T)|A|0(T)\rangle$, i.e., after diagonalizing $\mathcal H$.

Obviously, in most cases one cannot diagonalize $\mathcal H$
exactly and, thus, find the exact thermal vacuum and other
eigenstates. Usually, one resorts to some approximations and then
finds an approximate thermal vacuum state, e.g., the thermal vacua
corresponding to the HFB or the random phase approximations. In
the case that there appear several solutions in the given
approximation, one should find the minimum of thermodynamical
potential $\Omega$ to see which of them is realized. In TFD,
$\Omega$ is given as\cite{Umezawa75,Umezawa82}
\begin{equation}
\Omega = \langle \Psi_{0}(T) | H - \hat K T |\Psi_{0}(T)\rangle,
\end{equation}
where $\Psi_{0}(T)$ is the approximate thermal vacuum and $\hat K$
is the entropy operator of the system.

\section{The Finite Temperature QPM}\label{main}

\subsection{The thermal QPM Hamiltonian}

In what follows we will use the microscopic Hamiltonian of the
Quasiparticle-Phonon nuclear model\cite{Soloviev92}
 \begin{equation}\label{QPM_zero}
 H_{\rm QPM}=H_{\rm sp}+H_{\rm pair}+H_{\rm ph}.
 \end{equation}
The Hamiltonian (\ref{QPM_zero}) includes average fields of
protons and neutrons
\begin{equation}\label{H_sp}
 H_{\rm sp}=\sum_\tau{\sum_{jm}}^{\tau}(E_j-\lambda_\tau)a^\dag_{jm}a^{\phantom{\dag}}_{jm}~,
\end{equation}
pairing interactions of the BCS type
\begin{equation}\label{H_pair}
  H_{\rm pair}=-\frac14\sum_{\tau}G_{\tau}{\sum_{\genfrac{}{}{0pt}{1}{j_1m_1}{j_2m_2}}}^{\!\!\tau}
 a^\dag_{j_1m_1}a^\dag_{\overline{\jmath_1m_1}}
 a^{\phantom{\dag}}_{\overline{\jmath_2m_2}}a^{\phantom{\dag}}_{j_2m_2}\quad
 (a_{\overline{\jmath m}}=(-1)^{j-m}a_{ j-m})~,
\end{equation}
and effective  multipole-multipole isoscalar and isovector forces
\begin{equation}\label{H_ph}
 H_{\rm ph}=-\frac12\sum_{\lambda\mu}\sum_{\tau\rho=\pm1}
 (\kappa_0^{(\lambda)}+\rho\kappa_1^{(\lambda)})M^\dag_{\lambda\mu}(\tau)M^{\phantom{\dag}}_{\lambda\mu}(\rho\tau).
 \end{equation}
The single-particle operator $M^\dag_{\lambda\mu}(\tau)$ reads
 \begin{equation}\label{mult}
 M^\dag_{\lambda\mu}(\tau)={\sum_{\genfrac{}{}{0pt}{1}{j_1m_1}{j_2m_2}}}^{\!\!\tau}
 \langle j_1m_1|i^\lambda R_\lambda(r)Y_{\lambda\mu}(\theta,\phi)|j_2m_2\rangle
 a^\dag_{j_1m_1}a^{\phantom{\dag}}_{j_2m_2}.
 \end{equation}
Here, an operator $a^\dag_{jm}$  ($a^{\phantom{\dag}}_{jm}$) is  the creation (annihilation)  operator
of nucleon in a single-particle subshell with quantum numbers $nljm\equiv jm$ and energy $E_j$;
 index $\tau=n,p$ is the isotopic
one, changing the sign of $\tau$ means the interchange
$n\leftrightarrow p$, the notation ${\sum}^\tau$ implies a
summation over neutron ($\tau = n$) or proton ($\tau = p$)
single-particle states only. The parameter $G_\tau$ is the
proton-proton or neutron-neutron pairing interaction constant;
$R_\lambda(r)$ is the radial form factor of the $\lambda$-pole
separable interaction, $Y_{\lambda\mu}(\theta,\phi)$ is the
corresponding spherical harmonic, $\kappa_0^{(\lambda)}$ and
$\kappa_1^{(\lambda)}$ are the coupling constants of isoscalar and
isovector multipole-multipole interactions of $\lambda$
multipolarity. The value $\lambda_\tau$ is the neutron  or proton
chemical potential (the Fermi level).

To find an excitation spectrum of a hot nucleus  governed by the
QPM Hamiltonian (\ref{QPM_zero}), at the beginning we should
construct the thermal QPM Hamiltonian ${\cal H}_{\rm QPM}$
\begin{equation}\label{QPM_T}
{\cal H}_{\rm QPM}=H_{\rm QPM}-\widetilde H_{\rm QPM},
\end{equation}
where $\widetilde H_{\rm QPM}=\widetilde H_{\rm sp}+\widetilde H_{\rm pair}+\widetilde H_{\rm ph}$
is the tilde counterpart of $H_{\rm QPM}$ created by the tilde conjugation rules~(\ref{tc_rule}).
Then we should diagonalize $\mathcal{H}_{\rm QPM}$. This procedure is quite similar to that used in
the standard QPM,\cite{Soloviev92} i.e., at $T=0$. The main difference lies in the doubled Hilbert
space of a heated nucleus.

\subsection{Thermal quasiparticles}\label{qp_approx}

The first step is diagonalization of part of the full thermal
Hamiltonian, namely the sum of the two first terms ${\cal H}_{\rm
sp}+{\cal H}_{\rm pair}$ which in the following will be referred
to as the thermal BCS Hamiltonian $\mathcal H_{\rm BCS}$. To this
aim, we make the Bogoliubov $u, v$- transformation from nucleon
operators $a^\dag, a$ to quasiparticle  operators $\alpha^\dag,
\alpha$
\begin{eqnarray}\label{B_tr}
 \alpha^\dag_{jm}=u_{j}a^\dag_{jm}-v_{j}a_{\overline{\jmath m}}~,&\nonumber\\
 \alpha^{\phantom\dag}_{jm}=u_{j}a^{\phantom\dag}_{jm}-v_{j}{a}^\dag_{\overline{\jmath m}}~,
 &(u^2_j+v^2_j=1)~.
\end{eqnarray}
The same transformation with the same $u, v$ coefficients has to
be applied to nucleonic tilde operators $\widetilde
a^\dag_{jm},~\widetilde a^{\phantom{\dag}}_{jm}$, thus producing
the tilde quasiparticle operators $\widetilde\alpha^\dag_{jm}$ and
$\widetilde\alpha^{\phantom{\dag}}_{jm}$.

Thermal effects appear after the second (or thermal) Bogoliubov
transformation which mixes ordinary and tilde quasiparticle
operators and, thus, produces the operators of so-called thermal
quasiparticles $\beta^\dag_{jm}, \beta_{jm}$ and their tilde
counterparts
\begin{eqnarray}\label{T_tr}
  \beta^\dag_{jm}=x_j\alpha^\dag_{jm}-\sigma y_j\widetilde\alpha_{jm}~,& \nonumber\\
  \widetilde\beta^\dag_{jm}=x_j\widetilde\alpha^\dag_{jm}+\sigma y_j\alpha_{jm}~,&
  (x^2_j+y^2_j=1)~.
\end{eqnarray}
Note that in contrast with
Refs.\cite{Tana88,Hats89,Civi93,Kos94,Kos95} and many others we
include the factor $\sigma$ from the definition of DTCR
(\ref{DTC_rule}) to the thermal Bogoliubov transformation
(\ref{T_tr}).

To find the coefficients $u,\ v$, we express the thermal
Hamiltonian in terms of thermal quasiparticle operators
(\ref{T_tr}) and then require that the one-body part of the
thermal BCS Hamiltonian has to be diagonal in terms of thermal
quasiparticles. This leads to the following equations for the $u,\
v$ coefficients:
\begin{equation}\label{uv}
  u^2_j=\frac{1}{2}\left(1+\frac{E_j-\lambda_\tau}
   {\varepsilon_j}\right),\quad
   v^2_j=\frac{1}{2}\left(1-\frac{E_j-\lambda_\tau}
   {\varepsilon_j}\right)
\end{equation}
where $\varepsilon_j=\sqrt{(E_j-\lambda_{\tau})^2+\Delta^2_{\tau}}$. The gap parameter $\Delta_\tau$
and the chemical potential $\lambda_\tau$ are the solutions of the equations
\begin{equation}\label{BCS}
\Delta_\tau=\frac{G_\tau}{2}{\sum_j}^\tau(2j+1)(x^2_j-y^2_j)u_jv_j,\quad
N_\tau={\sum_j}^\tau(2j+1)(v^2_jx^2_j+u^2_jy^2_j),
\end{equation}
where $N_\tau$ is the number of neutrons or protons in a nucleus.
Actually, the equation for $N_\tau$ is not a consequence of the
procedure described above but rather an additional demand that a
number of particles of any kind in the heated system is conserved
in average.

With $u_j, v_j$ from (\ref{uv}) the one-body part of the thermal
BCS Hamiltonian reads
\begin{equation}\label{Ht_sp}
{\mathcal H}_{\rm BCS} = {\cal H}_{\rm sp}+{\cal H}_{\rm
pair}\simeq\sum_\tau{\sum_{jm}}^\tau\varepsilon_j
(\beta^\dag_{jm}\beta^{\phantom{\dag}}_{jm}-\widetilde\beta^\dag_{jm}\widetilde\beta^{\phantom{\dag}}_{jm}).
\end{equation}
Thus, $\mathcal H_{\rm BCS}$ describes a system of noninteracting
thermal quasiparticles and tilde-quasiparticles with energies
$\varepsilon_j$ and $-\varepsilon_j$, respectively. The vacuum for
thermal quasiparticles is given by\cite{Umezawa75,Ojima81}
\begin{equation}\label{tqp_vacuum}
 |0(\beta,\widetilde\beta)\rangle=
 \exp\left\{-\hat K_f\!/2\right\}\exp\Bigl\{\sigma^*\!\sum_\tau{\sum_{jm}}^{\tau}\!
 \alpha^{\dag}_{jm}\widetilde \alpha^{\dag}_{jm}\Bigr\}|0(\alpha)\rangle |0(\widetilde\alpha)\rangle~,
 \end{equation}
where $|0(\alpha)\rangle$ and $|0(\widetilde\alpha)\rangle$ is the
vacua for ordinary and tilde Bogoliubov quasiparticles,
respectively. The operator $\hat K_f$ is the entropy operator. It
reads
\begin{equation}
  \hat K_f=-\sum_\tau{\sum_{jm}}^\tau\{\alpha^\dag_{jm}\alpha^{\phantom{\dag}}_{jm}\ln y^2_j+
  \alpha^{\phantom{\dag}}_{jm}\alpha^\dag_{jm}\ln x^2_j\}~.
\end{equation}

We should stress that although the vacuum (\ref{tqp_vacuum}) is
the eigenstate of the thermal BCS Hamiltonian (\ref{Ht_sp}) with
zero eigenvalue it is not yet a thermal vacuum state in the sense
of (\ref{exp_val}). To determine the thermal vacuum state
corresponding to $\mathcal H_{\rm BCS}$ we need to fix
appropriately the coefficients $x_j, y_j$. They can be found by
minimizing the thermodynamic potential
\begin{eqnarray}\label{tdp_fermi}
 \Omega_f&=&
 \langle0(\beta,\widetilde\beta)|(H_{\rm sp}+H_{\rm pair})-T\hat
 K_f|0(\beta,\widetilde\beta)\rangle  \nonumber\\
  &=&\sum_\tau{\sum_{jm}}^\tau\bigl\{\varepsilon_jy^2_j+T(y^2_j\ln y^2_j+x^2_j\ln x^2_j)\bigr\}.
\end{eqnarray}
Note that $\Omega_f$ contains the ordinary operators $H_{\rm sp}$
and $H_{\rm pair}$ but not the thermal ones $\mathcal H_{\rm sp}$
and $\mathcal H_{\rm pair}$. As a result of variational procedure,
we obtain
\begin{eqnarray}\label{occup}
y_j=\left[1+\exp\left(\frac{\varepsilon_j}{T}\right)\right]^{-1/2}~,
  \quad x_j=\bigl(1-y^2_j\bigr)^{1/2}~.
\end{eqnarray}
Thus, the coefficients $y^2_j$ are nothing else than the thermal
occupation factors of the Fermi-Dirac statistics.

The thermal quasiparticle vacuum (\ref{tqp_vacuum}) with the
coefficients $y_j, x_j$ (\ref{occup}) is the thermal vacuum in the
thermal BCS approximation. Hereafter, it will be denoted by
$|0(T);{\rm qp}\rangle_\sigma$.

The average number of thermally excited  Bogoliubov quasiparticles
with quantum numbers $jm$ in the BCS thermal vacuum $|0(T);{\rm
qp}\rangle_\sigma$ is
\begin{equation}
  {}_\sigma\langle 0(T);{\rm qp}|
  \alpha^\dag_{jm}\alpha^{\phantom{\dag}}_{jm}
  |0(T);{\rm qp}\rangle_\sigma=y^2_{j},
\end{equation}
whereas the average number of nucleons in the same state is
\begin{equation}
  n_j(T)={}_\sigma\langle 0(T);{\rm qp}|
  a^\dag_{jm}a^{\phantom{\dag}}_{jm}
  |0(T);{\rm qp}\rangle_\sigma=u^2_jy^2_{j}+v^2_jx^2_j~.
\end{equation}
The function $n_j(T)$ determines smearing of the Fermi surface due
to thermal and pairing effects.

Equations (\ref{BCS}) with $y_j, x_j$ (\ref{occup})  are the
well-known BCS-equations at finite temperature (see
e.g.\cite{Good81,Civi83,Ign83}).

Since the thermal vacuum $|0(T);{\rm qp}\rangle_\sigma$ contains a
certain number of  Bogoliubov quasiparticles,  the excited states
at finite temperature can be built on the top of $|0(T);{\rm
qp}\rangle_\sigma$ by either adding or eliminating a Bogoliubov
quasiparticle. Due to the relations
\begin{equation}
\alpha^\dag_{jm}|0(T);{\rm
qp}\rangle_\sigma\!=\!x_j\beta^\dag_{jm}|0(T);{\rm
qp}\rangle_\sigma,\quad \alpha^{\phantom\dag}_{\overline{\jmath
m}}|0(T);{\rm qp}
\rangle_\sigma\!=\!\sigma^*y_j\widetilde\beta^\dag_{\overline{\jmath
m}}|0(T);{\rm qp}\rangle_\sigma
\end{equation}
one can associate
the first process  with creation of a thermal quasiparticle having
a positive energy, whereas  the second process can be considered
as creation of a tilde thermal quasiparticle having a negative
energy.

The simplest excitations on the top of the BCS thermal vacuum in an even-even hot nucleus involve
two thermal quasiparticles. Their wave functions and energies are
 \begin{eqnarray}\label{thermal_tq}
  \bigl[\beta^\dag_{j_1}\beta^\dag_{j_2}\bigr]^\lambda_\mu|0(T);{\rm qp}\rangle_\sigma,~~
  \omega\!&=&\!\varepsilon_{j_1}\!+\!\varepsilon_{j_2}\equiv\varepsilon^{(+)}_{j_1j_2};\nonumber\\
  \bigl[\widetilde\beta^\dag_{\overline{\jmath_1}}
        \widetilde\beta^\dag_{\overline{\jmath_2}}\bigr]^\lambda_\mu|0(T);{\rm qp}\rangle_\sigma,~~
  \omega\!&=&\!-\varepsilon^{(+)}_{j_1j_2};\nonumber\\
  \bigl[\beta^\dag_{j_1}\widetilde\beta^\dag_{\overline{\jmath_2}}\bigr]^\lambda_\mu|0(T);{\rm
  qp}\rangle_\sigma,~~
 \omega\!&=&\!\varepsilon_{j_1}\!-\!\varepsilon_{j_2}\equiv\varepsilon^{(-)}_{j_1j_2};\nonumber\\
   \bigl[\widetilde\beta^\dag_{\overline{\jmath_1}}\beta^\dag_{j_2}\bigr]^\lambda_\mu|0(T);{\rm
   qp}\rangle_\sigma,~~
 \omega\!&=&\!-\varepsilon^{(-)}_{j_1j_2}.
  \end{eqnarray}
The square brackets  $[~~]^\lambda_\mu$ in (\ref{thermal_tq}) mean
the coupling of single-particle momenta $j_1$, $j_2$ to the total
angular momentum $\lambda$ with the magnetic quantum number $\mu$.

As it should be, any thermal two-quasiparticle state with positive
energy $\omega$ has a counterpart -- a tilde-conjugated state with
negative energy $-\omega$.

Quite interesting relations exist between electromagnetic
transition probabilities to thermal two-quasiparticle states and
their tilde counterparts. Let us write the $E\lambda$-transition
operator $\mathcal{M}(E\lambda\mu)$ in terms of  thermal
quasiparticles
\begin{eqnarray}\label{el_tr_oper}
\mathcal{M}(E\lambda\mu)&=&\frac{1}{\hat\lambda}\sum_{\tau}{\sum_{j_1j_2}}^{\tau}
\Gamma^{(\lambda)}_{j_1j_2}
\left\{A^\dag_{\lambda\mu}(j_1j_2)+A_{\overline{\lambda\mu}}(j_1j_2)+B_{\lambda\mu}(j_1j_2)\right\},
\nonumber \\
 A^\dag_{\lambda\mu}(j_1j_2)&=&\frac12
 u^{(+)}_{j_1j_2}\Bigl(x_{\!j_1}x_{\!j_2}[\beta^\dag_{j_1}\beta^\dag_{j_2}]^\lambda_\mu\!-\!
 \sigma^2y_{\!j_1}y_{\!j_2}[\widetilde{\beta}^\dag_{\overline{\jmath_1}}
 \widetilde{\beta}^\dag_{\overline{\jmath_2}}]^\lambda_\mu\Bigr)\!-\!\sigma^*
 v^{(-)}_{j_1j_2}x_{\!j_1}y_{\!j_2}[\beta^\dag_{j_1}\widetilde{\beta}^\dag_{\overline{\jmath_2}}]^\lambda_\mu~,
 \nonumber\\
A_{\overline{\lambda\mu}}(j_1j_2)&=&
(-1)^{\lambda-\mu}\left[A^\dag_{\lambda-\mu}(j_1j_2)\right]^\dag,\nonumber\\
 B_{\lambda\mu}(j_1j_2)&=&
 -v^{(-)}_{j_1j_2}\Bigl(x_{\!j_1}x_{\!j_2}[\beta^\dag_{j_1}\beta^{\phantom{\dag}}_{\overline{\jmath_2}}]^\lambda_\mu+
 y_{\!j_1}y_{\!j_2}[\widetilde{\beta}^\dag_{\overline{\jmath_1}}
 \widetilde{\beta}^{\phantom{\dag}}_{j_2}]^\lambda_\mu\Bigr)\nonumber\\&&
 \quad\quad\quad\quad\quad
 +u^{(+)}_{j_1j_2}\Bigl(\sigma{x}_{\!j_1}y_{\!j_2}[\beta^\dag_{j_1}\widetilde{\beta}^{\phantom{\dag}}_{j_2}]^\lambda_\mu-
 \sigma^*y_{\!j_1}x_{\!j_2}[\widetilde{\beta}^\dag_{\overline{\jmath_1}}
 \beta^{\phantom{\dag}}_{\overline{\jmath_2}}]^\lambda_\mu\Bigr).
\end{eqnarray}
In (\ref{el_tr_oper}), $\Gamma^{(\lambda)}_{j_1j_2}$ is the
reduced single-particle matrix element of the
$E\lambda$-transition operator;
$u^{(+)}_{j_1j_2}=u_{j_1}v_{j_2}+u_{j_2}v_{j_1}$,
$v^{(-)}_{j_1j_2}=u_{j_1}u_{j_2}-v_{j_1}v_{j_2}$ and
$\hat\lambda=\sqrt{2\lambda+1}$.

Only the terms containing the operators
$A^\dag_{\lambda\mu}(j_1j_2)$ and
$A_{\overline{\lambda\mu}}(j_1j_2)$ contribute to transitions from
the BCS thermal vacuum state to any thermal two-quasiparticle
state. Introducing the functions
\begin{equation}
Y(\omega)=\left[\exp\left(\frac{\omega}{T}\right)-1\right]^{-1/2};
\quad X(\omega)=[1 + Y^2(\omega)]^{1/2}
\end{equation}
and  taking the advantage of the relations
 \begin{eqnarray}\label{xyXY}
 y^2_{j_1}y^2_{j_1}&=&(1-y^2_{j_1}-y^2_{j_2})
 Y^2(\varepsilon^{(+)}_{j_1j_2}), \nonumber\\
  x^2_{j_1}y^2_{j_2}&=&(y^2_{j_2}-y^2_{j_1})
 Y^2(\varepsilon^{(-)}_{j_1j_2}),~
  ~({\rm for}~~\varepsilon_{j_1}\!>\varepsilon_{j_2})
 \end{eqnarray}
we get the squared reduced matrix elements $\Phi^2_{\lambda}$ of
the operator $\mathcal{M}(E\lambda\mu)$ between the BCS thermal
vacuum and different thermal two-quasiparticle states
(\ref{thermal_tq})
  \begin{eqnarray}\label{prob1}
   \Phi^2_\lambda([\beta^\dag_{j_1}\beta^\dag_{j_2}]^\lambda_\mu)&=&
  \bigl(\Gamma^{(\lambda)}_{j_1j_2}u^{(+)}_{j_1j_2}\bigr)^2(1-y^2_{j_1}-y^2_{j_2})
  X^2(\varepsilon^{(+)}_{j_1j_2}),
  \nonumber\\ [4mm]
 \Phi^2_\lambda([\widetilde\beta^\dag_{\overline{\jmath_1}}
 \widetilde\beta^\dag_{\overline{\jmath_2}}]^\lambda_\mu)&=&
\bigl(\Gamma^{(\lambda)}_{j_1j_2}u^{(+)}_{j_1j_2}\bigr)^2(1-y^2_{j_1}-y^2_{j_2})Y^2(\varepsilon^{(+)}_{j_1j_2}),
  \nonumber\\ \rule{0mm}{10mm}
 \Phi^2_\lambda([\beta^\dag_{j_1}\widetilde\beta^\dag_{\overline{\jmath_2}}]^\lambda_\mu)&=&
  \bigl(\Gamma^{(\lambda)}_{j_1j_2}v^{(-)}_{j_1j_2}\bigr)^2 \times
   \left\{\begin{array}{ll}
   (y^2_{j_2}-y^2_{j_1})X^2(\varepsilon^{(-)}_{j_1j_2}),&~(\varepsilon_{j_1}\!>\!\varepsilon_{j_2}),\\
   (y^2_{j_1}-y^2_{j_2})Y^2(\varepsilon^{(-)}_{j_2j_1}),&~(\varepsilon_{j_1}\!<\!\varepsilon_{j_2}),
   \rule{0pt}{15pt}\end{array}\right.\nonumber
      \\   \rule{0mm}{10mm}
 \Phi^2_\lambda([\widetilde\beta^\dag_{\overline{\jmath_1}}\beta^\dag_{j_2}]^\lambda_\mu)&=&
 \bigl(\Gamma^{(\lambda)}_{j_1j_2}v^{(-)}_{j_1j_2}\bigr)^2 \times
    \left\{\begin{array}{ll}
     (y^2_{j_2}-y^2_{j_1})Y^2(\varepsilon^{(-)}_{j_1j_2}),&~(\varepsilon_{j_1}\!>\!\varepsilon_{j_2}),\\
     (y^2_{j_1}-y^2_{j_2})X^2(\varepsilon^{(-)}_{j_2j_1}),&~(\varepsilon_{j_1}\!<\!\varepsilon_{j_2}),
      \rule{0pt}{15pt}
     \end{array}\right.
 \end{eqnarray}
From (\ref{prob1}) it follows that transition probabilities to the
states tilde-conjugated to each other and, correspondingly, having
the energies $\pm \omega$ relate with the factor
 \begin{equation}\label{symm_qp}
 \Phi^2_{\lambda}(\omega)=\exp\left(\frac{\omega}{T}\right)\Phi^2_{\lambda}(-\omega).
 \end{equation}
Formally, the function $Y(\omega)$ in (\ref{prob1}) is the
Bose-Einstein distribution function which determines the average
number of bosons with energy $\omega$ in a system in the thermal
equilibrium at temperature $T$. This gives an idea to treat
two-fermion excitations in a hot nucleus as bosons. The
probability to create a boson is proportional to the factor
$[1+Y^2(\omega)]$, whereas the probability to annihilate it is
proportional to $Y^2(\omega)$.

Expressions (\ref{prob1})  determine $E\lambda$-strength
distribution in a hot nucleus within the independent BCS
quasiparticle approximation. An interesting point is that in
contrast with $T=0$ case a portion of the $E\lambda$-strength
appears in the negative energy region $\omega<0$, i.e., below the
thermal vacuum state, at finite temperatures. This strength
determines a probability of $\gamma$-ray emission by a hot
nucleus, whereas the strength at $\omega>0$ determines a
photoabsorption cross section. Both the parts of the
$E\lambda$-strength contribute to the energy weighted sum rule
(EWSR) at $T\neq0$
\begin{multline}\label{EWSR_qp}
\text{EWSR} = \sum_\tau{\sum_{j_1\ge j_2}}^\tau\varepsilon^{(+)}_{j_1j_2}\left[
 \Phi^2_\lambda([\beta^\dag_{j_1}\beta^\dag_{j_2}]^\lambda_\mu)-
 \Phi^2_\lambda([\widetilde\beta^\dag_{\overline{\jmath_1}}
 \widetilde\beta^\dag_{\overline{\jmath_2}}]^\lambda_\mu) \right]\\
 +\sum_\tau{\sum_{j_1\ge j_2}}^\tau\varepsilon^{(-)}_{j_1j_2}\left[
 \Phi^2_\lambda([\beta^\dag_{j_1}\widetilde\beta^\dag_{\overline{\jmath_2}}]^\lambda_\mu)-
 \Phi^2_\lambda([\widetilde\beta^\dag_{\overline{\jmath_1}}
 \beta^\dag_{j_2}]^\lambda_\mu) \right]\\
=\sum_\tau{\sum_{j_1\ge j_2}}^\tau\bigl(\Gamma^{(\lambda)}_{j_1j_2}\bigr)^2
\left[\varepsilon^{(+)}_{j_1j_2}(u^{(+)}_{j_1j_2})^2(1-y^2_{j_1}-y^2_{j_2})-
      \varepsilon^{(-)}_{j_1j_2}(v^{(-)}_{j_1j_2})^2(y^2_{j_1}-y^2_{j_2})\right].
\end{multline}

\subsection{Thermal phonons}\label{RPA_approx}

The second step in the diagonalization of the thermal Hamiltonian
is to take into account the long-range particle-hole interaction
$H_{\rm ph}$. This interaction is responsible for the existence of
different types of collective vibrations in nuclei. In this
Subsection, the thermal quasiparticle random phase approximation
in treating the vibrations of hot nuclei is discussed.

Transformations (\ref{B_tr}) and  (\ref{T_tr}) with the
coefficients determined by Eqs.(\ref{uv}) and (\ref{occup}) should
be applied to the rest of the thermal Hamiltonian
 $H_{\rm ph}- \widetilde{H}_{\rm ph}$ as well. Then the thermal Hamiltonian (\ref{QPM_T}) takes the form
\begin{multline}\label{QPM_TFD}
{\mathcal H}_{\rm QPM}=\sum_\tau{\sum_{jm}}^{\tau}\varepsilon_j
 (\beta^\dag_{jm}\beta^{\phantom{\dag}}_{jm}-
 \widetilde\beta^\dag_{jm}\widetilde\beta^{\phantom{\dag}}_{jm})\\-
 \frac12\sum_{\lambda\mu}\sum_{\tau\rho=\pm1}
 (\kappa_0^{(\lambda)}+\rho\kappa_1^{(\lambda)})\left\{M^\dag_{\lambda\mu}(\tau)M^{\phantom{\dag}}_{\lambda\mu}(\rho\tau)
 -\widetilde M^\dag_{\lambda\mu}(\tau)\widetilde
 M^{\phantom{\dag}}_{\lambda\mu}(\rho\tau)\right\}.
\end{multline}
In terms of thermal quasiparticles the multipole operator $M^\dag_{\lambda\mu}(\tau)$ has the same
shape as the $E\lambda$-transition operator (\ref{el_tr_oper}). The only difference is that one
should substitute matrix elements $\Gamma^{(\lambda)}_{j_1j_2}$ for matrix elements
$f^{(\lambda)}_{j_1j_2}=\langle j_1\|i^\lambda R_\lambda(r)Y_{\lambda}(\theta,\phi)\|j_2\rangle$.

The thermal Hamiltonian (\ref{QPM_TFD}) can be approximately
reduced to the Hamiltonian of noninteracting bosonic excitations
-- thermal phonons. This occurs if one omits in (\ref{QPM_TFD})
the terms containing the operator $B_{\lambda\mu}(j_1j_2)$.
Hereafter, the remaining part of the thermal Hamiltonian
(\ref{QPM_TFD}) is denoted  ${\mathcal H}_{\rm TRPA}$.

To diagonalize ${\mathcal H}_{\rm TRPA}$, it is natural to use a
trial wave function which is a linear superposition of different
types of thermal two-quasiparticle operators, namely
\begin{multline}\label{phonon}
  Q^\dag_{\lambda\mu i}\!=\!\frac12\sum_\tau{\sum_{j_1j_2}}^\tau
 \Bigl(\psi^{\lambda i}_{j_1j_2}[\beta^\dag_{j_1}\beta^\dag_{j_2}]^\lambda_\mu\!\!+
 \widetilde\psi^{\lambda i}_{j_1j_2}[\widetilde\beta^\dag_{\overline{\jmath_1}}
 \widetilde\beta^\dag_{\overline{\jmath_2}}]^\lambda_\mu\!\!+
 2\sigma\eta^{\lambda i}_{j_1j_2}[\beta^\dag_{j_1}
  \widetilde\beta^\dag_{\overline{\jmath_2}}]^\lambda_\mu\Bigr)\\
+(-1)^{\lambda-\mu}\!\left(
 \phi^{\lambda i}_{j_1j_2}[\beta_{j_1}\beta_{j_2}]^\lambda_{-\mu}\!\!+
 \widetilde\phi^{\lambda i}_{j_1j_2}[\widetilde\beta_{\overline{\jmath_1}}
 \widetilde\beta_{\overline{\jmath_2}}]^\lambda_{-\mu}\!\!+
 2\sigma^*\xi^{\lambda i}_{j_1j_2}[\beta_{j_1}
  \widetilde\beta_{\overline{\jmath_2}}]^\lambda_{-\mu}\!\right).
\end{multline}
The factors $\sigma$ and $\sigma^*$ at the cross-over (i.e.,
tilde-nontilde) terms of (\ref{phonon}) are absent in the thermal
phonon definition in.\cite{Kos94,Kos95,Storo04} They appear due to
adoption of the new DTCR~(\ref{DTC_rule}). The definition
(\ref{phonon}) coincides with the one in\cite{Kos94,Kos95,Storo04}
when $\sigma=1$.

One more very important assumption has to be accepted. We assume
that the thermal biquasiparticle operators contained in
(\ref{phonon}) commute like bosonic operators
  \begin{align}\label{QBA}
\Bigl[[\beta^{\phantom{\dag}}_{j_1}\beta^{\phantom{\dag}}_{j_2}]^\lambda_\mu~,
      [\beta^\dag_{j_3}\beta^\dag_{j_4}]^{\lambda'}_{\mu'}\Bigr]&\approx-
      \delta_{\lambda\lambda'}\delta_{\mu\mu'}\Bigl(
      \delta_{j_1\!j_3}\delta_{j_2\!j_4}+
      (-1)^{j_1-j_2+\lambda}\delta_{j_1\!j_4}\delta_{j_2\!j_3}\Bigr),
      \nonumber\\[2mm]
\Bigl[[\beta^{\phantom{\dag}}_{j_1}\widetilde\beta^{\phantom{\dag}}_{j_2}]^\lambda_\mu~,
      [\beta^\dag_{j_3}\widetilde\beta^\dag_{j_4}]^{\lambda'}_{\mu'}\Bigr]&\approx-
      \delta_{\lambda\lambda'}\delta_{\mu\mu'}
      \delta_{j_1j_3}\delta_{j_2j_4}.
  \end{align}
and all other commutators are equal to zero. This assumption is
known as the quasiboson approximation.

Moreover, taking into consideration the long-range interaction we should redefine the thermal
vacuum state. At first, we define it as the vacuum state $|0(Q,\widetilde Q)\rangle_\sigma$ for
thermal phonons
\begin{equation}
             Q_{\lambda\mu i}|0(Q,\widetilde Q)\rangle_\sigma=0,\quad
  \widetilde Q_{\lambda\mu i}|0(Q,\widetilde Q)\rangle_\sigma=0~.
\end{equation}
A thermal one-phonon state is constructed by acting on the thermal
phonon vacuum $|0(Q,\widetilde Q)\rangle_\sigma$ by the thermal
phonon creation operator
\begin{equation}\label{one_ph_st}
          Q^\dag_{\lambda\mu i}|0(Q,\widetilde Q)\rangle_\sigma .
\end{equation}
A thermal tilde one-phonon state should be define as
\begin{equation}\label{til_one_ph_st}
 \widetilde Q^\dag_{\overline{\lambda\mu} i}|0(Q,\widetilde Q)\rangle_\sigma\equiv
 (-1)^{\lambda-\mu}
 \widetilde Q^\dag_{\lambda-\mu i}|0(Q,\widetilde Q)\rangle_\sigma
\end{equation}
because just the operator  $(-1)^{\lambda-\mu}\widetilde Q^\dag_{\lambda-\mu i}$ transforms  under
spatial rotations like a spherical tensor of rank $\lambda$.

The set of thermal one-phonon wave functions has to be
orthonormalized. This demand together with the quasiboson
approximation for commutators of biquasiparticle operators
(\ref{QBA}) imposes the following constraints on the phonon
amplitudes $\psi, \phi, \widetilde{\psi}, \widetilde{\phi}, \eta,
\xi$:
\begin{eqnarray}\label{constr}
 \frac12
 \sum_{\tau}{\sum_{j_1j_2}}^\tau
  g^{\lambda i}_{j_1j_2}w^{\lambda i'}_{j_1j_2}+
 \widetilde{g}^{\lambda i}_{j_1j_2}\widetilde{w}^{\lambda i'}_{j_1j_2}+
 t^{\lambda i}_{j_1j_2}s^{\lambda i'}_{j_1j_2}+
 \widetilde{t}^{\lambda i}_{j_1j_2}\widetilde{s}^{\lambda i'}_{j_1j_2}&=&\delta_{ii'},\nonumber\\
 \sum_{\tau}{\sum_{j_1j_2}}^\tau
  g^{\lambda i}_{j_1j_2}\widetilde w^{\lambda i'}_{j_1j_2}+
 \widetilde{g}^{\lambda i}_{j_1j_2}{w}^{\lambda i'}_{j_1j_2}+
 t^{\lambda i}_{j_1j_2}\widetilde s^{\lambda i'}_{j_1j_2}+
 \widetilde t^{\lambda i}_{j_1j_2}s^{\lambda i'}_{j_1j_2}&=&0.
 \end{eqnarray}
In (\ref{constr}), the following notation is introduced for the
sums and differences of original phonon amplitudes:
 \begin{align}
 \binom{g}{w}^{\lambda i}_{j_1j_2}&=
 \psi^{\lambda i}_{j_1j_2}\pm\phi^{\lambda i}_{j_1j_2},
 & \binom{\widetilde g}{\widetilde w}^{\lambda i}_{j_1j_2}&=
 \widetilde\psi^{\lambda i}_{j_1j_2}\pm\widetilde\phi^{\lambda i}_{j_1j_2},\nonumber\\[3mm]
\binom{t}{s}^{\lambda i}_{j_1j_2}&=
 \eta^{\lambda i}_{j_1j_2}\pm\xi^{\lambda i}_{j_1j_2},
 & \binom{\widetilde t}{\widetilde s}^{\lambda i}_{j_1j_2}&=
 \widetilde\eta^{\lambda i}_{j_1j_2}\pm\widetilde \xi^{\lambda i}_{j_1j_2}.
\end{align}
Moreover, $\widetilde\eta^{\lambda i}_{j_1j_2}\equiv(-1)^{j_1-j_2+\lambda}\ \eta^{\lambda
i}_{j_2j_1}$, $\widetilde\xi^{\lambda i}_{j_1j_2}\equiv(-1)^{j_1-j_2+\lambda}\ \xi^{\lambda
i}_{j_2j_1}$.

Constraints (\ref{constr}) imply that thermal phonon operators
commute like bosons.

With constraints (\ref{constr}) one can find the inverse
transformation to (\ref{phonon})
\begin{eqnarray}\label{inverse}
 [\beta^\dag_{j_1}\beta^\dag_{j_2}]^\lambda_\mu&=&\phantom{\sigma^*}{\sum_i}^\tau
 \psi^{\lambda i}_{j_1j_2}Q^\dag_{\lambda\mu i}+
 \phi^{\lambda i}_{j_1j_2}Q_{\overline{\lambda\mu} i}+
 \widetilde\psi^{\lambda i}_{j_1j_2}\widetilde Q^\dag_{\overline{\lambda\mu} i}+
 \widetilde\phi^{\lambda i}_{j_1j_2}\widetilde Q_{\lambda\mu i}~,\nonumber\\
 \bigl[\beta^\dag_{j_1}\widetilde\beta^\dag_{\overline{\jmath_2}}\bigr]^\lambda_\mu&=&\sigma^*{\sum_i}^\tau
 \eta^{\lambda i}_{j_1j_2}Q^\dag_{\lambda\mu i}+
 \xi^{\lambda i}_{j_1j_2}Q_{\overline{\lambda\mu}i}+
 \widetilde\eta^{\lambda i}_{j_1j_2}\widetilde Q^\dag_{\overline{\lambda\mu} i}+
 \widetilde\xi^{\lambda i}_{j_1j_2}\widetilde Q_{\lambda\mu i} ,
  \end{eqnarray}
and then evaluate the following expression of the thermal RPA Hamiltonian $\mathcal H_{\rm TRPA}$
in terms of thermal phonon operators:
\begin{multline}\label{H_RPA2}
{\cal H}_{\rm TRPA}=\sum_\tau{\sum_{jm}}^{\tau}\varepsilon_j
 \bigl(\beta^\dag_{jm}\beta^{\phantom{\dag}}_{jm}-
 \widetilde\beta^\dag_{jm}\widetilde\beta^{\phantom{\dag}}_{jm}\bigr)\\
 -\frac{1}{8}\sum_{\lambda\mu i i'}\!\frac{1}{\hat\lambda^2}\!\!
 \sum_{\tau\rho=\pm1}\!\!(\kappa^{\lambda}_0+\rho\kappa^{\lambda}_1)\!
 \left\{\!\bigl[D^{\lambda i}_\tau D^{\lambda i'}_{\rho\tau}\!-\!
  \widetilde D^{\lambda i}_\tau\widetilde D^{\lambda i'}_{\rho\tau}\bigr]\!
 \bigl(Q^\dag_{\lambda\mu i}\!+\!Q_{\overline{\lambda\mu}i}\bigr)\!
 \bigl(Q^\dag_{\overline{\lambda\mu}i'}\!+\!Q^{\phantom\dag}_{\lambda\mu i'}\bigr)\right.\\
 \qquad+\left.\bigl[D^{\lambda i}_\tau\widetilde D^{\lambda i'}_{\rho\tau}\!-\!
  \widetilde D^{\lambda i}_\tau D^{\lambda i'}_{\rho\tau}\bigr]\!
 \bigl(Q^\dag_{\lambda\mu i}\!+\!Q_{\overline{\lambda\mu}i}\bigr)\!
 \bigl(\widetilde Q^\dag_{\lambda\mu i'}\!+
 \!\widetilde Q_{\overline{\lambda\mu}i'}\bigr)-({\rm
 t.c.})\right\}.
 \end{multline}
The notation "(t.c.)" in (\ref{H_RPA2}) stands for the items which
are tilde conjugated to the displayed ones. The functions
$D^{\lambda i}_{\tau}$ and $\widetilde D^{\lambda i}_{\tau}$
($\tau=n,p$) are the following combinations of phonon amplitudes:
\begin{eqnarray}\label{D_tau}
 D^{\lambda i}_{\tau}\!&=&\!{\sum_{j_1j_2}}^\tau f^{(\lambda)}_{j_1j_2}
 \bigl[u^{(+)}_{j_1j_2}(x_{j_1}x_{j_2}g^{\lambda i}_{j_1j_2}-
 \sigma^2y_{j_1}y_{j_2}\widetilde g^{\lambda i}_{j_1j_2})-
 2\sigma^2v^{(-)}_{j_1j_2}x_{j_1}y_{j_2}t^{\lambda
 i}_{j_1j_2}\bigr],\nonumber\\
 \widetilde D^{\lambda i}_{\tau}\!&=\!&{\sum_{j_1j_2}}^\tau f^{(\lambda)}_{j_1j_2}
 \bigl[u^{(+)}_{j_1j_2}(x_{j_1}x_{j_2}\widetilde{g}^{\lambda i}_{j_1j_2}-
 \sigma^2y_{j_1}y_{j_2} g^{\lambda i}_{j_1j_2})-
 2\sigma^2v^{(-)}_{j_1j_2}y_{j_1}x_{j_2}t^{\lambda
 i}_{j_1j_2}\bigr].
 \end{eqnarray}

To find eigenvalues of ${\cal H}_{\rm TRPA}$, we apply the
variational principle, i.e., we minimize the expectation value of
${\mathcal H}_{\rm TRPA}$ over the thermal one-phonon state under
constraints (\ref{constr}).

It should be stressed that the phonon vacuum $|0(Q,\widetilde
Q)\rangle_\sigma$ is not the thermal vacuum in the sense of
Eq.~(\ref{exp_val}) and, thus, the expectation value of any
physical operator with respect to $|0(Q,\widetilde
Q)\rangle_\sigma$ does not correspond to the average over the
grand canonical ensemble.

After a variation procedure with respect to functions $g$, $w$,
$\widetilde{g}$, $\widetilde{w}$, $t$, and $s$ one gets a
homogeneous system of linear equations. Since we use the separable
effective interaction, the system of equations looks simple
\begin{eqnarray}\label{RPA_eq}
\binom{\psi}{\phi}^{\lambda i}_{\tau j_1j_2}&=& \frac{1}{2\hat\lambda^2}
\frac{f^{(\lambda)}_{j_1j_2}u^{(+)}_{j_1j_2}} {\varepsilon_{j_1j_2}^{(+)}\mp\omega_{\lambda
i}}\sum_{\rho=\pm1}
 (\kappa^{(\lambda)}_0+\rho\kappa^{(\lambda)}_1)
  \bigl[x_{j_1}x_{j_2}D^{\lambda i}_{\rho\tau}+
\sigma^2y_{j_1}y_{j_2}\widetilde D^{\lambda i}_{\rho\tau}\bigr]~,\nonumber\\
 \binom{\widetilde\psi}{\widetilde\phi}^{\lambda i}_{\tau j_1j_2}&=&
 \frac{\sigma^2}{2\hat\lambda^2}
\frac{f^{(\lambda)}_{j_1j_2}u^{(+)}_{j_1j_2}} {\varepsilon_{j_1j_2}^{(+)}\pm\omega_{\lambda
i}}\sum_{\rho=\pm1}
 (\kappa^{(\lambda)}_0+\rho\kappa^{(\lambda)}_1)
  \bigl[y_{j_1}y_{j_2} D^{\lambda i}_{\rho\tau}+
        \sigma^2x_{j_1}x_{j_2}\widetilde D^{\lambda i}_{\rho\tau}\bigr]~,\nonumber\\
\binom{\eta}{\xi}^{\lambda i}_{\tau j_1j_2}&=&- \frac{\sigma^2}{2\hat\lambda^2}
\frac{f^{(\lambda)}_{j_1j_2}v^{(-)}_{j_1j_2}} {\varepsilon_{j_1j_2}^{(-)}\!\mp\omega_{\lambda
i}}\sum_{\rho=\pm1}
 (\kappa^{(\lambda)}_0+\rho\kappa^{(\lambda)}_1)
  \bigl[x_{j_1}y_{j_2}D^{\lambda i}_{\rho\tau}-
y_{j_1}x_{j_2}\widetilde D^{\lambda i}_{\rho\tau}\bigr].
\end{eqnarray}

For a further discussion it is more convenient to rewrite the
system (\ref{RPA_eq}) regarding the functions $D^{\lambda
i}_{\rho\tau}$ and $\widetilde D^{\lambda i}_{\rho\tau}$ as
unknown variables. Then we get
\begin{equation}\label{RPA_eq2}
\sum_{\rho=\pm1}(\kappa^{(\lambda)}_0+\rho\kappa^{(\lambda)}_1)\binom{D}{\widetilde
D}^{\lambda i}_{\rho\tau}=
\bigl[X^{(\lambda)}_\tau(\omega_{\lambda
i})\bigr]^{-1}\binom{D}{\widetilde D}^{\lambda i}_\tau,
\end{equation}
where the function $X^{\lambda}_\tau(\omega)$ reads
 \begin{eqnarray}
\!\!\!\!\!\!
  X^{(\lambda)}_\tau(\omega)\!=\!\frac{1}{\hat\lambda^2}{\sum_{j_1j_2}}^\tau
 (f^{(\lambda)}_{\!j_1j_2})^2\!\left[
 \frac{(u^{(+)}_{j_1j_2})^2\varepsilon_{\!j_1j_2}^{(+)}(1\!-y^2_{\!j_1}\!-y^2_{\!j_2})}
       {(\varepsilon_{\!j_1j_2}^{(+)})^2-\omega^2}\!-\!
 \frac{(v^{(-)}_{\!j_1j_2})^2\varepsilon_{j_1j_2}^{(-)}(y^2_{\!j_1}\!-y^2_{\!j_2})}
       {(\varepsilon_{\!j_1j_2}^{(-)})^2-\omega^2}\right].
  \end{eqnarray}

Demanding the existence of a nontrivial solution to the system of
linear equations (\ref{RPA_eq2}), we derive the secular equation
for the energy of the thermal one-phonon state $\omega_{\lambda
i}$
 \begin{equation}\label{secular}
 (\kappa^{(\lambda)}_0+\kappa^{(\lambda)}_1)
 \Bigl[X^{(\lambda)}_p(\omega)+X^{(\lambda)}_{n}(\omega)\Bigr]-
 4\kappa^{(\lambda)}_0\kappa^{(\lambda)}_1X^{(\lambda)}_p(\omega)X^{(\lambda)}_{n}(\omega)=1.
 \end{equation}
Equation (\ref{secular}) is strictly the same as was obtained by
other methods in.\cite{Ignat75,Som83,Alas90} It does not differ as
well from the previous results of the TFD-QPM
approach.\cite{Kos94,Kos95}

One obtains the same secular equation as (\ref{secular}) if the
variational procedure is applied to the expectation value of
${\cal H}_{\rm TRPA}$ over the tilde thermal one-phonon state
$\widetilde Q^\dag_{\overline{\lambda\mu} i}|0(Q,\widetilde
Q)\rangle_\sigma$. As it was stated in the previous subsection,
both the positive and negative energy excitations of a hot nucleus
built on the top of the thermal vacuum  have a physical meaning
and should be considered on equal footing. So the negative roots
of (\ref{secular}) are identified with energies of tilde-phonons.

Unfortunately, the values $D^{\lambda i}_{\tau}$ and $\widetilde D^{\lambda i}_{\tau}$ cannot be
determined unambiguously from equations (\ref{RPA_eq2}) and the normalization condition
(\ref{constr}). The reason for this indeterminacy is the following. Equations (\ref{RPA_eq2})
enable one to prove that the TRPA Hamiltonian ${\cal H}_{\rm TRPA}$ is diagonal in terms of the
thermal phonon operators, i.e.,
\begin{equation}\label{H_TRPA}
 {\cal H}_{\rm TRPA}=\sum_{\lambda\mu
i}\omega_{\lambda i}\bigl(Q^{\dag}_{\lambda\mu i}Q^{\phantom{\dag}}_{\lambda\mu i}- \widetilde
Q^{\dag}_{\lambda\mu i}\widetilde Q^{\phantom{\dag}}_{\lambda\mu i}\bigr)
\end{equation}
It is seen that the Hamiltonian (\ref{H_TRPA}) is invariant under
the unitary transformation which mixes non-tilde and tilde thermal
phonons but preserves the secular equation (\ref{secular}). Just
due to this invariance the systems of equations (\ref{RPA_eq}) or
(\ref{RPA_eq2}) cannot be solved unambiguously. To overcome this
problem, one needs to involve additional considerations. To this
aim, minimization of the thermodynamic potential was used while
considering the pairing correlations at finite temperature (see
Subsec.(\ref{RPA_approx})). However, in the case of a thermal
phonon system this procedure is not so straightforward.

Let us turn back to equations (\ref{RPA_eq}). The following interrelations between $D^{\lambda
i}_{\tau}$ and $\widetilde D^{\lambda i}_{\tau}$ can be easily found from the normalization
condition~\eqref{constr}:
\begin{equation}\label{D_norma}
\bigl(D^{\lambda i}_{\tau}\bigr)^2-\bigr(\widetilde D^{\lambda
i}_{\tau}\bigl)^2=4\hat\lambda^4\frac{
X^{(\lambda)}_\tau(\omega_{\lambda i})}{{\cal N}^{\lambda
i}_{\tau}}~,
\end{equation}
where ${\cal N}^{\lambda i}_{\tau}$ is given by
\begin{multline}
 {\cal N}^{\lambda i}_\tau=\hat\lambda^2\left[
 \frac{\partial}{\partial\omega}X^{(\lambda)}_\tau(\omega)\Bigl|_{\omega=\omega_{\lambda
 i}}\right.\\ \left.+
 \left(\frac{1-X^{(\lambda)}_\tau(\omega_{\lambda i})(\kappa^{(\lambda)}_0+\kappa^{(\lambda)}_1)}
 {X^{(\lambda)}_{-\tau}(\omega_{\lambda i})(\kappa^{(\lambda)}_0-\kappa^{(\lambda)}_1)}\right)^2
 \frac{\partial}{\partial\omega}X^{(\lambda)}_{-\tau}(\omega)\Bigl|_{\omega=\omega_{\lambda i}}
 \right].
\end{multline}

Instead of the functions $D^{\lambda i}_{\tau}$ and $\widetilde
D^{\lambda i}_{\tau}$ it is convenient to introduce the new
variables
\begin{equation}
X_{\lambda i}= \frac{D^{\lambda i}_\tau\sqrt{{\cal N}^{\lambda
i}_\tau}} {2\hat\lambda^2X^{(\lambda)}_\tau(\omega_{\lambda
i})}~,\quad
Y_{\lambda i}= \frac{\widetilde D^{\lambda i}_\tau\sqrt{{\cal
N}^{\lambda i}_\tau}}
{2\hat\lambda^2X^{(\lambda)}_\tau(\omega_{\lambda i})},
\end{equation}
which obey the following condition: $ X^2_{\lambda i}-Y^2_{\lambda i}=1$.

Then we substitute $X_{\lambda i}$ and $Y_{\lambda i}$ for
$D^{\lambda i}_{\tau}$ and $\widetilde D^{\lambda i}_{\tau}$ in
equations (\ref{RPA_eq}) and get
\begin{eqnarray}\label{amplitudes}
 \binom{\psi}{\phi}^{\lambda i}_{j_1j_2}&=&
 \frac{1}{\sqrt{{\cal N}^{\lambda i}_\tau}}
 \frac{f^{(\lambda)}_{j_1j_2}u^{(+)}_{j_1j_2}}
      {\varepsilon^{(+)}_{j_1j_2}\mp\omega_{\lambda i}}
      \bigl(x_{j_1}x_{j_2}X_{\lambda i}+\sigma^2y_{j_1}y_{j_2}Y_{\lambda i}\bigr)~,\nonumber\\
 \rule{0mm}{10mm}
 \binom{\widetilde\psi}{\widetilde\phi}^{\lambda i}_{ j_1j_2}&=&
 \frac{\sigma^2}{\sqrt{{\cal N}^{\lambda i}_\tau}}
 \frac{f^{(\lambda)}_{j_1j_2}u^{(+)}_{j_1j_2}}
      {\varepsilon^{(+)}_{j_1j_2}\pm\omega_{\lambda i}}
      \bigl(y_{j_1}y_{j_2} X_{\lambda i}+\sigma^2x_{j_1}x_{j_2} Y_{\lambda i}\bigr)~, \nonumber\\
 \rule{0mm}{10mm}
 \binom{\eta}{\xi}^{\lambda i}_{j_1j_2}&=&-
 \frac{\sigma^2}{\sqrt{{\cal N}^{\lambda i}_\tau}}
 \frac{f^{(\lambda)}_{j_1j_2}v^{(-)}_{j_1j_2}}
      {\varepsilon^{(-)}_{j_1j_2}\mp\omega_{\lambda i}}
      \bigl(x_{j_1}y_{j_2}X_{\lambda i} -y_{j_1}x_{j_2}Y_{\lambda
      i}\bigr).
\end{eqnarray}

Some conclusions concerning variables $X_{\lambda i}$ and
$Y_{\lambda i}$ can be achieved if one analyzes their behaviour at
the limit $T\to0$. Obviously, the phonon wave function
(\ref{phonon}) at $T=0$ should consist of two components only:
$\alpha^{\dag}_{j_1}\alpha^{\dag}_{j_2}$ and
$\alpha_{j_2}\alpha_{j_1}$. Since at $T\to0$ the thermal
occupation numbers of quasiparticles $y_j$ tend to zero ($x_j
\to1$), the demand can be fulfilled only if $Y_{\lambda i}\to0$
(and synchronously $X_{\lambda i}\to1$). In such a case the
tilde-amplitudes ${\widetilde\psi}^{\lambda i}_{ j_1j_2}$ and
${\widetilde\phi}^{\lambda i}_{ j_1j_2}$ tend to zero whereas
$\psi^{\lambda i}_{ j_1j_2}$ and $\phi^{\lambda i}_{ j_1j_2}$
survive.

Let us define the thermal phonons with the amplitudes
corresponding to $Y_{\lambda i}=0$ and $X_{\lambda i}=1$ as
$q^{\dag}_{\lambda \mu i},\ q_{\lambda \mu i}$ and name them
"reference phonons" (r-phonons). The vacuum for r-phonons is
denoted by $|0(q,\widetilde q)\rangle_\sigma$. From
(\ref{amplitudes}) one can conclude that the thermal phonons
corresponding to other values of $X_{\lambda i},\ Y_{\lambda i}$
can be produced by applying the  unitary $\{X_{\lambda i},\
Y_{\lambda i}\}$ transformation to the r-phonons
\begin{eqnarray}\label{bos_therm_rot}
    Q_{\lambda\mu i}^\dag &=&
    X_{\lambda i}q_{\lambda\mu i}^\dag-Y_{\lambda i}\widetilde{q}_{\lambda\mu i}^{\phantom{\dag}},
      \nonumber  \\
    \widetilde{Q}_{\lambda\mu i}^\dag &=& X_{\lambda i}\widetilde{q}_{\lambda\mu i}^\dag-
    Y_{\lambda i}q_{\lambda\mu i}^{\phantom{\dag}}.
\end{eqnarray}
The vacuum for the phonons $Q_{\lambda\mu i}^\dag,\ \widetilde Q_{\lambda\mu i}^\dag$ has the
form\cite{Umezawa75}
\begin{equation}\label{tph_vacuum}
 |0(Q,\widetilde Q)\rangle_\sigma=\exp\left\{-\hat K_b/2\right\}\exp\Bigl\{\sum_{\lambda\mu i}
  q^{\dag}_{\lambda\mu i}\widetilde q^{\dag}_{\lambda\mu i}\Bigr\}|0(q,\widetilde
  q)\rangle_\sigma,
 \end{equation}
where $\hat K_b$ is the entropy operator for noninteracting bosons
\begin{equation}
  \hat K_b=-\sum_{\lambda\mu i}\{q^\dag_{\lambda\mu i}q^{\phantom{\dag}}_{\lambda\mu i}\ln Y^2_{\lambda i}-
  q^{\phantom{\dag}}_{\lambda\mu i}q^\dag_{\lambda\mu i}\ln X^2_{\lambda i}\}.
\end{equation}

Now we are ready to determine the values $X_{\lambda i},\
Y_{\lambda i}$. To this aim, we minimize the thermodynamic
potential $\Omega_{b}$ for the r-phonon system which reads
 \begin{eqnarray}\label{B_T_P}
 \Omega_{b}&=&{}_\sigma\langle0(Q,\widetilde Q)|
 \sum_{\lambda\mu i}\omega_{\lambda i}q^\dag_{\lambda\mu i}q^{\phantom{\dag}}_{\lambda\mu i}
  -T\hat K_b|0(Q,\widetilde Q)\rangle_\sigma\nonumber\\&=&
\sum_{\lambda\mu i}\left\{\omega_{\lambda i}Y^2_{\lambda i}+T(Y^2_{\lambda i}\ln Y^2_{\lambda i}+
X^2_{\lambda i}\ln X^2_{\lambda i})\right\}.
\end{eqnarray}
Varying (\ref{B_T_P}) with respect to $Y_{\lambda i}$ and equating
the result to zero we get
\begin{equation}\label{ph_oc_n}
  Y_{\lambda i}=\left[\exp\left(\frac{\omega_{\lambda
  i}}{T}\right)-1\right]^{-1/2},\qquad
  X_{\lambda i}=\bigl[1+Y_{\lambda i}^2\bigr]^{1/2}.
\end{equation}
The coefficients  $Y^2_{\lambda i}$ are the thermal occupation
factors of the Bose-Einstein statistics. They determine the
average number of r-phonons in the thermal phonon vacuum
\begin{equation}\label{BE}
{}_\sigma\langle0(Q,\widetilde Q)|\ q^\dag_{\lambda\mu i}q^{\phantom{\dag}}_{\lambda\mu i}|\
0(Q,\widetilde Q)\rangle_\sigma=Y^2_{\lambda
  i}~.
\end{equation}

It is worthwhile to note that the problems of correct construction of the thermal RPA phonon
operator and the thermal RPA vacuum state were already discussed in\cite{Civi93} using the BCS
Hamiltonian as an example. The authors of\cite{Civi93} also confronted with an ambiguity of RPA
solutions at finite temperature. To get it out, they introduced the "non-rotated" thermal pairing
phonons and then made a thermal rotation which minimized the thermodynamic potential. At the same
time, these circumstances, i.e., the invariance of the $\mathcal H_{\rm TRPA}$ under the thermal
rotation (\ref{bos_therm_rot}), were overlooked in.\cite{Kos94,Kos95} In these papers, it was
assumed that $Y_{\lambda i}=0$ at any temperature and the vacuum for "r-phonons" was identified as
the "true" thermal vacuum state. That is why  thermal bosonic occupation numbers did not appear in
the corresponding formulae in.\cite{Kos94,Kos95,Storo04}

Now we can fix the factor $\sigma$. To this aim, we consider the
behaviour of thermal phonon amplitudes (\ref{amplitudes}) when the
coupling constants of a separable multipole interaction
$\kappa_{0,1}^{(\lambda)}$ tend to zero. When
$\kappa_{0,1}^{(\lambda)}\to 0$ the thermal phonon energy
$\omega_{\lambda i}\to \varepsilon^{(\pm)}_{j_{1}j_{2}}$ and  only
one amplitude has to survive in the corresponding phonon wave
function. In particular, all the amplitudes of backward going
components vanish when $\kappa_{0,1}^{(\lambda)}\to 0$. Taking
advantage of (\ref{xyXY}) it can be shown that if $\omega_{\lambda
i}\to\varepsilon^{(-)}_{j_1j_2}$ $\xi^{\lambda i}_{j_1j_2}\to 0$
whereas $(\eta^{\lambda i}_{j_1j_2})^2\to1$ at any $\sigma$ value.

When $\omega_{\lambda i}\to\varepsilon^{(+)}_{j_1j_2}$ the
amplitudes $\phi$ and $\widetilde \psi$ also vanish at any value
of $\sigma$. However, limiting values of $\psi$ and $\widetilde
\phi$ appear to be dependent on $\sigma$, namely,
\begin{eqnarray}
\!\!\!\!\lim_{\omega_{\lambda i}\to\varepsilon^{(+)}_{j_1 j_2}} \psi^{\lambda i}_{j_1j_2} =
 \frac{x_{j_1}x_{j_2}X(\varepsilon^{(+)}_{j_1j_2})
 +\sigma^2y_{j_1}y_{j_2}Y(\varepsilon^{(+)}_{j_1j_2})}
 {(1-y_{j_1}^2-y_{j_2}^2)^{1/2}}=
 X^2(\varepsilon^{(+)}_{j_1j_2})+\sigma^2Y^2(\varepsilon^{(+)}_{j_1j_2})
  ,\quad\nonumber\\
\lim_{\omega_{\lambda i}\to\varepsilon^{(+)}_{j_1 j_2}}\!\widetilde\phi^{\lambda i}_{j_1j_2}\!=\!
 \frac{\sigma^2y_{j_1}y_{j_2}X(\varepsilon^{(+)}_{j_1j_2})\!+\!
 x_{j_1}x_{j_2}Y(\varepsilon^{(+)}_{j_1j_2})}
 {(1-y_{j_1}^2-y_{j_2}^2)^{1/2}}=
 X(\varepsilon^{(+)}_{j_1j_2})Y(\varepsilon^{(+)}_{j_1j_2})(\sigma^2\!+\!1).\qquad
\end{eqnarray}
Since from a physical point of view $\widetilde\phi^{\lambda i}_{j_1j_2}$ should vanish, we choose
$\sigma=i$. Then not only $\widetilde\phi^{\lambda i}_{j_1j_2}=0$ but also $\psi^{\lambda
i}_{j_1j_2} = 1$. The present result on DTCR agrees with the conclusion in.\cite{Mats85}
In,\cite{Mats85} the appropriate choice of DTCR was specified by the fermion number conservation in
the system. If the number of fermions in the system is not conserved, DTCR should have the form
(\ref{Ojima_dtcr}), i.e. $\widetilde{\widetilde{A}}=A$. This seems to be just our case since the
number of quasiparticles in a nucleus is not conserved.

Now we complete constructing  the thermal phonon operator and the "true" thermal vacuum state in
the random phase approximation. They should be calculated with formulae (\ref{phonon}) and
(\ref{amplitudes}) where $X_{\lambda i}$ and $Y_{\lambda i}$ are given by (\ref{ph_oc_n}) and
$\sigma = i$. Hereafter, the thermal phonon vacuum is denoted by $|0(T);{\rm TRPA}\rangle$. We
would like to stress once again that the thermal RPA vacuum  $|0(T);{\rm TRPA}\rangle$ reduces to
the thermal BCS vacuum when the particle-hole interaction vanishes only if the coefficients
$X_{\lambda i}$, $Y_{\lambda i}$ are the phonon thermal occupation numbers given by (\ref{ph_oc_n})
and $\sigma=i$.

At the end of this subsection we calculate a matrix element of the $E\lambda$- transition between
the RPA thermal vacuum and a thermal one-phonon state. To this aim, one has to write the operator
${\mathcal M}(E\lambda\mu)$ (\ref{el_tr_oper}) in terms of thermal phonon operators. Taking into
account only the term of (\ref{el_tr_oper}) with the operators $A^\dag_{\lambda\mu}(j_1j_2)$ and
$A_{\overline{\lambda\mu}}(j_1j_2)$ one gets
\begin{multline}\label{El_operator}
 {\cal M}(E\lambda\mu) =
 \frac{1}{2\hat\lambda}\sum_i\sum_\tau{\sum_{j_1j_2}}^\tau\Gamma^{(\lambda)}_{j_1j_2}\\
   \times\Bigl\{\bigl[u^{(+)}_{j_1j_2}
   (x_{j_1}x_{j_2}g^{\lambda i}_{j_1j_2}+
    y_{j_1}y_{j_2}\widetilde g^{\lambda i}_{j_1j_2})+
    2v^{(-)}_{j_1j_2}x_{j_1}y_{j_2}t^{\lambda i}_{j_1j_2}\bigr]
    \bigl(Q^\dag_{\lambda\mu i}+Q_{\overline{\lambda\mu}i}\bigr)\\
  ~~~~~+\bigl[u^{(+)}_{j_1j_2}
   (x_{j_1}x_{j_2}\widetilde g^{\lambda i}_{j_1j_2}+
    y_{j_1}y_{j_2} g^{\lambda i}_{j_1j_2})+
    2v^{(-)}_{j_1j_2}y_{j_1}x_{j_2}t^{\lambda i}_{j_1j_2}\bigr]
    \bigl(\widetilde Q^\dag_{\overline{\lambda\mu}i}+\widetilde Q_{\lambda\mu i})\Bigr\}\\
 =\frac{1}{\hat\lambda}\sum_i\sum_\tau
  \Gamma^{\lambda i}_\tau\left\{
   X_{\lambda i}\bigl(Q^\dag_{\lambda\mu i}+Q_{\overline{\lambda\mu}i}\bigr)+
   Y_{\lambda i}\bigl(\widetilde Q^\dag_{\overline{\lambda\mu}i}+\widetilde Q_{\lambda\mu
   i}\bigr)\right\},
\end{multline}
where
\begin{eqnarray}
\Gamma^{\lambda i}_\tau={\sum_{j_1j_2}}^\tau \
  \frac{\Gamma^{(\lambda)}_{j_1j_2}f^{(\lambda)}_{j_1j_2}}
       {\sqrt{{\cal N}^{\lambda i}_\tau}}\left[
 \frac{(u^{(+)}_{j_1j_2})^2\varepsilon_{\!j_1j_2}^{(+)}(1\!-y^2_{\!j_1}\!-y^2_{\!j_2})}
       {(\varepsilon_{\!j_1j_2}^{(+)})^2-\omega^2_{\lambda i}}-
 \frac{(v^{(-)}_{\!j_1j_2})^2\varepsilon_{j_1j_2}^{(-)}(y^2_{\!j_1}\!-y^2_{\!j_2})}
       {(\varepsilon_{\!j_1j_2}^{(-)})^2-\omega^2_{\lambda i}}\right].~~
 \end{eqnarray}
Expression (\ref{El_operator}) differs significantly from that
in.\cite{Kos95,Storo04} Specifically, there appear terms with
tilde-phonon operators in (\ref{El_operator}). This is a
consequence of $\{X, Y \}$ rotation of thermal phonons. The item
proportional to the factor $Y_{\lambda i}$  is responsible for
transitions to tilde-phonon states lying below the thermal vacuum
state, i.e., for decay of the thermal vacuum.

Thus, there are two types of matrix elements corresponding to
excitation and de-excitation processes of the thermal vacuum
\begin{eqnarray}\label{ampl1}
    \Phi_{\lambda i}&=&
    \langle0(T);{\rm TRPA}\|{\cal M}(E\lambda\mu)Q^{\dag}_{\lambda\mu i}\|0(T);{\rm TRPA}\rangle=
    X_{\lambda i}\sum_\tau \Gamma^{\lambda i}_\tau,\nonumber
    \\
    \widetilde\Phi_{\lambda i}&=&
    \langle0(T);{\rm TRPA}\|{\cal M}(E\lambda\mu)
    \widetilde{Q}^{\dag}_{\overline{\lambda\mu}i}\|0(T);{\rm TRPA}\rangle=
    Y_{\lambda i}\sum_\tau \Gamma^{\lambda i}_\tau.
 \end{eqnarray}
The factors $X_ {\lambda i}$ and $Y_{\lambda i}$ in (\ref{ampl1})
were missed in\cite{Kos95,Storo04} since the r-phonons
$q^{\dag}_{\lambda \mu i}, \widetilde q^{\dag}_{\lambda \mu i}$
were explored rather than the "rotated" $Q^{\dag}_{\lambda \mu i},
\widetilde Q^{\dag}_{\lambda \mu i}$.

The factor $X_{\lambda i}^2$ in the photoabsorption cross section for hot nuclei as well as its
role were discussed in detail in.\cite{Chom90} This factor occurred also in the response function
in\cite{Som83} but was missed in the $B(E\lambda)\downarrow$ expression in.\cite{Alas90} Note that
in both the papers\cite{Som83,Alas90} the thermal RPA equations were derived with the equation of
motion method for bifermion operators $\alpha^{\dag}_{j_{1}}\alpha^{\dag}_{j_{2}}$ and
$\alpha^{\dag}_{j_{1}}\alpha_{j_{2}}$.

As in the case of transitions to  thermal two-quasiparticle states
(see Subsec.(\ref{qp_approx})), there exists the following
relation between transition probabilities from the non-tilde and
tilde one-phonon states
\begin{equation}\label{symmetry}
\Phi^2_{\lambda i}=\exp(\omega_{\lambda i}/T)\widetilde\Phi^2_{\lambda i}.
\end{equation}
This relation is equivalent to the principle of detailed balancing
connecting the probabilities for the probe to transfer energy
$\omega$ to the heated system and to absorb energy $\omega$ from
the heated system (see e.g.\cite{Som83}).

The model energy-weighed sum rule at the thermal RP approximation
is given by
\begin{equation}\label{EWSR_ph}
 \mathrm{EWSR}=
 \sum_{\omega_{\lambda i}>0}\omega_{\lambda i}\bigl(\Phi_{\lambda i}^2-\widetilde\Phi_{\lambda i}^2\bigr)=
 \sum_{\omega_{\lambda i}>0}\omega_{\lambda i}
 \left(\Gamma^{\lambda i}_p+\Gamma^{\lambda i}_n\right)^2.
\end{equation}
It is worth-while to note that the EWSR (\ref{EWSR_ph}) appears to
be independent on the thermal phonon occupation factors. The
numerical value of (\ref{EWSR_ph}) should coincide with that of
(\ref{EWSR_qp}).

Nominally, this expression coincides with that in.\cite{Alas90} However, the essential difference
between the present result and\cite{Alas90} is the contribution of TRPA states with negative
energies $\omega_{\lambda i}$ (i.e., tilde states). In,\cite{Alas90} the total strength of
$E\lambda$-transitions is located in the positive energy region whereas in the present approach
some fraction of strength is carried by the states with $\omega_{\lambda i}<0$. As a result, part
of the EWSR pertaining to positive excitation energies is greater than the total one that is
obvious from (\ref{EWSR_ph}).

\subsection{Interaction of thermal phonons}\label{high_approx}

In this Subsection, we deal with a coupling of thermal phonons,
i.e., go beyond the thermal RPA. Physical effects which can be
treated in this order of approximation relate to fragmentation of
basic nuclear excitations like quasiparticles and phonons, their
spreading widths and/or more consistent description of transition
strength distributions over a nuclear spectrum. The problem of
temperature dependence of the giant resonance width which was
intensively discussed not long ago (see, e.g.,
reviews\cite{Egido93,DiToro00,Shlomo05}) also belongs to this set.

The part of $\mathcal H$ (\ref{QPM_T}) which is responsible for
these effects is the so-called quasiparticle-phonon interaction
(or the cubic anharmonic term) ${\cal H}_{\rm qph}$
\begin{eqnarray}\label{th_Ham3}
 &&{\cal H}_{\rm qph}=-
 \frac12\sum_{\lambda\mu i}\sum_\tau\!{\sum_{j_1j_2}}^\tau
 \frac{f^{(\lambda)}_{j_1j_2}}{\sqrt{{\cal N}^{\lambda i}_\tau}}\left\{
 \bigl(Q^\dag_{\overline{\lambda\mu}i}+Q^{\phantom{\dag}}_{\lambda\mu i}\bigr)
  B_{\lambda\mu i}(j_1j_2)+({\rm h.c.})-({\rm t.c.})\right\},
  \nonumber \\ && \nonumber \\
  &&B_{\lambda\mu i}(j_1j_2)=iu^{(+)}_{j_1j_2}\left(
  {\cal Z}_{j_1j_2}^{\lambda i}[\beta^\dag_{j_1}\widetilde{\beta}^{\phantom{\dag}}_{j_2}]^\lambda_\mu+
  {\cal Z}_{j_2j_1}^{\lambda i}[\widetilde{\beta}^\dag_{\overline{\jmath_1}}
                         \beta^{\phantom{\dag}}_{\overline{\jmath_2}}]^\lambda_\mu\right)
  \nonumber\\&&\qquad\qquad\qquad\qquad\qquad\qquad
  -v^{(-)}_{j_1j_2}\left(
  {\cal X}_{j_1j_2}^{\lambda i}[\beta^\dag_{j_1}\beta^{\phantom{\dag}}_{\overline{\jmath_2}}]^\lambda_\mu+
  {\cal Y}_{j_1j_2}^{\lambda i}[\widetilde{\beta}^\dag_{\overline{\jmath_1}}
                         \widetilde{\beta}^{\phantom{\dag}}_{j_2}]^{\lambda}_{\mu}\right).
   \end{eqnarray}
The coefficients ${\cal X}_{j_1j_2}^{\lambda i}$, ${\cal
Y}_{j_1j_2}^{\lambda i}$ and ${\cal Z}_{j_1j_2}^{\lambda i}$ are
given by
\begin{equation}
 \binom{\cal X}{\cal Y}_{j_1j_2}^{\lambda i}=x_{j_1}x_{j_2}\binom{X}{Y}_{\lambda i}+
                                   y_{j_1}y_{j_2}\binom{Y}{X}_{\lambda i}~,~~~
 {\cal Z}_{j_1j_2}^{\lambda i}=x_{j_1}y_{j_2}X_{\lambda i}+y_{j_1}x_{j_2}Y_{\lambda i}~.
 \end{equation}
The interaction ${\cal H}_{\rm qph}$ couples the multiphonon
states whose structures differ by one phonon, i.e., one-phonon
states with two-phonon states, two-phonon states with three-phonon
states etc. This term can be named the "cubic anharmonic" one
because in the lowest order of boson expansion the fermionic
operator $B_{\lambda\mu i}(j_1j_2)$ is substituted by the product
of two bosonic operators (or phonons). The remaining part of the
thermal Hamiltonian consists of the items $\sim
B^\dag_{\lambda-\mu}(j_1j_2)
B^{\phantom{\dag}}_{\lambda\mu}(j_3j_4)$ which are equivalent to
the sum of products of four bosonic operators. Its contribution
will not be considered here.

To take into account the term ${\cal H}_{\rm qph}$, we again apply
the variational principle. A new trial wave function is assumed to
be of the form
\begin{multline}\label{trial}
 |\Psi_\nu(JM)\rangle=\biggl\{\sum_i R_i(J\nu)Q^{\dag}_{JMi}+
 \sum_{\stackrel{\lambda_1i_1}{\lambda_2i_2}}P^{\lambda_1i_1}_{\lambda_2i_2}(J\nu)
 \bigl[Q^\dag_{\lambda_1i_1}Q^\dag_{\lambda_2i_2}\bigr]^J_M\\
 +\!\sum_{\stackrel{\lambda_1i_1}{\lambda_2 i_2}}\!S^{\lambda_1i_1}_{\lambda_2i_2}(J\nu)
 \bigl[Q^\dag_{\lambda_1i_1}\!\widetilde{Q}^\dag_{\overline{\lambda_2}i_2}\bigr]^J_M\!+\!
 \sum_{\stackrel{\lambda_1i_1}{\lambda_2 i_2}}\!T^{\lambda_1i_1}_{\lambda_2i_2}(J\nu)
 \bigl[\widetilde{Q}^\dag_{\overline{\lambda_1}i_1}\!\widetilde{Q}^\dag_{\overline{\lambda_2}i_2}\bigr]^J_M
 \biggr\}|0(T);{\rm RPA}\rangle,
\end{multline}
where $R,~P,~S,~T$ are the variational parameters which should be
determined. As one can see in (\ref{trial}), the thermal vacuum is
kept the same as in the TRPA. At $T=0$ the latter approximation is
valid if the quasiparticle-phonon interaction is relatively weak.

The trial wave function~(\ref{trial}) has to be normalized and,
therefore, the amplitudes $R,\ P,\ S,\ T$ should satisfy the
following constraints:
\begin{equation}\label{constr2}
 \sum_i\bigl[R_i(J\nu)\bigr]^2+
 \sum_{\stackrel{\lambda_1i_1}{\lambda_2i_2}}\left\{2\bigl[P^{\lambda_1i_1}_{\lambda_2i_2}(J\nu)\bigr]^2+
  \bigl[S^{\lambda_1i_1}_{\lambda_2i_2}(J\nu)\bigr]^2+
  2\bigl[T^{\lambda_1i_1}_{\lambda_2i_2}(J\nu)\bigr]^2\right\}=1.
\end{equation}

Since the trial wave function contains three different types of
components, there are three types of interaction matrix elements
which couple a thermal one-phonon state with two-phonon ones
\begin{eqnarray}
  U^{\lambda_1i_1}_{\lambda_2i_2}(Ji)&=&
  \langle{\rm RPA};0(T)|Q_{JMi}{\cal H}_{\rm qph}
  \bigl[Q^\dag_{\lambda_1i_1}Q^\dag_{\lambda_2i_2}\bigr]^J_M|0(T);{\rm RPA}\rangle,
  \nonumber\\&&\nonumber\\
  V^{\lambda_1i_1}_{\lambda_2i_2}(Ji)&=&
  \langle{\rm RPA};0(T)|Q_{JMi}{\cal H}_{\rm qph}
  \bigl[Q^\dag_{\lambda_1i_1}
  \widetilde{Q}^\dag_{\overline{\lambda_2}i_2}\bigr]^J_M|0(T);{\rm RPA}\rangle,
  \nonumber\\&&\nonumber \\
  W^{\lambda_1i_1}_{\lambda_2i_2}(Ji)&=&
  \langle{\rm RPA};0(T)|Q_{JMi}{\cal H}_{\rm qph}
  \bigl[\widetilde Q^\dag_{\overline{\lambda_1}i_1}
        \widetilde{Q}^\dag_{\overline{\lambda_2}i_2}\bigr]^J_M|0(T);{\rm
        RPA}\rangle.
\end{eqnarray}
The matrix element $U^{\lambda_1i_1}_{\lambda_2i_2}(Ji)$ correspond to a transition from one to two
phonons, whereas the matrix elements $V^{\lambda_1i_1}_{\lambda_2i_2}(Ji)$ and
$W^{\lambda_1i_1}_{\lambda_2i_2}(Ji)$ describe a thermal phonon scattering and a thermal phonon
absorption, respectively. The explicit forms of $U, V$ and $W$ are given in Appendix~A.

Thus, we should minimize the expectation value of the thermal
Hamiltonian $\mathcal{H}_{\rm RPA}+ \mathcal{H}_{\rm qph}$ over
$|\Psi_\nu(JM)\rangle$ at constraint (\ref{constr2}). The
expectation value is given by
\begin{multline}\label{QHQ2}
  \langle\Psi_\nu(JM)|\mathcal{H}_{\rm RPA}+ \mathcal{H}_{\rm qph}|\Psi_\nu(JM)\rangle=
  \sum_i\omega_{Ji}\bigl[R_i(J\nu)\bigr]^2\\
  +2\sum_{\stackrel{\lambda_1i_1}{\lambda_2i_2}}
  (\omega_{\lambda_1i_1}+\omega_{\lambda_2i_2})\bigl[P^{\lambda_1i_1}_{\lambda_2i_2}(J\nu)\bigr]^2
  +\sum_{\stackrel{\lambda_1i_1}{\lambda_2i_2}}
  (\omega_{\lambda_1i_1}-\omega_{\lambda_2i_2})\bigl[S^{\lambda_1i_1}_{\lambda_2i_2}(J\nu)\bigr]^2
   ~~\qquad\\
  -2\sum_{\stackrel{\lambda_1i_1}{\lambda_2i_2}}
  (\omega_{\lambda_1i_1}+\omega_{\lambda_2i_2})\bigl[T^{\lambda_1i_1}_{\lambda_2i_2}(J\nu)\bigr]^2
  +2\sum_i\sum_{\stackrel{\lambda_1i_1}{\lambda_2i_2}}\!R_i(J\nu)\left\{
  P^{\lambda_1i_1}_{\lambda_2i_2}(J\nu)U^{\lambda_1i_1}_{\lambda_2i_2}(Ji)\right.\\
  +\left.
  S^{\lambda_1i_1}_{\lambda_2i_2}(J\nu)W^{\lambda_1i_1}_{\lambda_2i_2}(Ji)+
  T^{\lambda_1i_1}_{\lambda_2i_2}(J\nu)V^{\lambda_1i_1}_{\lambda_2i_2}(Ji)\right\}.
\end{multline}

After the standard variation procedure one gets a homogeneous
system of linear equations ($\eta_{\nu}$ is the energy of the
state $|\Psi_\nu(JM)\rangle$)
\begin{eqnarray}\label{system2}
  R_i(J\nu)(\omega_{\lambda i}-\eta_\nu)
  + \sum_{\stackrel{\lambda_1i_1}{\lambda_2i_2}}\left\{
  P^{\lambda_1i_1}_{\lambda_2i_2}(J\nu)U^{\lambda_1i_1}_{\lambda_2i_2}(Ji) \right.
  \qquad\qquad\qquad&&\nonumber\\
  +\left.
  S^{\lambda_1i_1}_{\lambda_2i_2}(J\nu)V^{\lambda_1i_1}_{\lambda_2i_2}(Ji)+
  T^{\lambda_1i_1}_{\lambda_2i_2}(J\nu)W^{\lambda_1i_1}_{\lambda_2i_2}(Ji)\right\}&=&0~,\nonumber\\
  P^{\lambda_1i_1}_{\lambda_2i_2}(J\nu)(\omega_{\lambda_1i_1}+\omega_{\lambda_2i_2}-\eta_\nu)+
  \frac12\sum_i  R_i(J\nu)U^{\lambda_1i_1}_{\lambda_2i_2}(Ji)&=&0~,\nonumber\\
  S^{\lambda_1i_1}_{\lambda_2i_2}(J\nu)(\omega_{\lambda_1i_1}-\omega_{\lambda_2i_2}-\eta_\nu)+
 \phantom{\frac12} \sum_i  R_i(J\nu)V^{\lambda_1i_1}_{\lambda_2i_2}(Ji)&=&0~,\nonumber\\
  T^{\lambda_1i_1}_{\lambda_2i_2}(J\nu)(\omega_{\lambda_1i_1}+\omega_{\lambda_2i_2}+\eta_\nu)-
  \frac12\sum_i  R_i(J\nu)W^{\lambda_1i_1}_{\lambda_2i_2}(Ji)&=&0~.
\end{eqnarray}
The system has a solution if $\eta_\nu$ is the root of the
following secular equation:
\begin{multline}\label{det}
 {\rm
 det}\left|\bigl(\omega_{Ji}-\eta_\nu\bigr)\delta_{ii'}-\frac12\sum_{\stackrel{\lambda_1i_1}{\lambda_2i_2}}
 \left\{
 \frac{U^{\lambda_1i_1}_{\lambda_2i_2}(Ji)U^{\lambda_1i_1}_{\lambda_2i_2}(Ji')}
 {\omega_{\lambda_1i_1}+\omega_{\lambda_2i_2}-\eta_\nu} \right.\right.\\+\left.\left.
 2\frac{V^{\lambda_1i_1}_{\lambda_2i_2}(Ji)V^{\lambda_1i_1}_{\lambda_2i_2}(Ji')}
 {\omega_{\lambda_1i_1}-\omega_{\lambda_2i_2}-\eta_\nu}-
  \frac{W^{\lambda_1i_1}_{\lambda_2i_2}(Ji)W^{\lambda_1i_1}_{\lambda_2i_2}(Ji')}
  {\omega_{\lambda_1i_1}+\omega_{\lambda_2i_2}+\eta_\nu}\right\}\right|=0~.
\end{multline}
Minimizing the expectation value of the thermal Hamiltonian
$\mathcal{H}_{\rm RPA}+ \mathcal{H}_{\rm qph}$ over the  wave
function $|\widetilde{\Psi}_\nu(JM)\rangle$ we get the
corresponding equations for tilde-conjugated states. They also can
be obtained from (\ref{system2}) and (\ref{det}) by changing the
sign of $\eta_\nu$. That is if $\eta_\nu$ is the energy of a given
state $|\Psi_\nu (JM)\rangle$, then $-\eta_\nu$ is the energy of
the tilde-conjugated state $|\widetilde{\Psi}_\nu (JM)\rangle$.

Certainly, the above results deviate from those of.\cite{Kos94}
Due to the changes of thermal RPA-phonon amplitudes and, in
particular, their dependence on the thermal phonon occupation
numbers, the quasiparticle-phonon interaction (\ref{th_Ham3})
couples one-phonon states with three types of thermal two-phonon
states whereas in\cite{Kos94} only one type was taken into account
(namely, the first term in (\ref{trial}).

The terms $Q^\dag\widetilde{Q}^\dag\ {\rm and}\
\widetilde{Q}^\dag\widetilde{Q}^\dag$ in the trial wave function
describe the processes which were not considered in\cite{Kos94}
and later in.\cite{Storo04} The processes of thermal phonon
scattering and absorption became possible due to the presence of
thermal r-phonons in the thermal phonon vacuum. Since
in\cite{Kos94} the thermal quasiparticle-phonon interaction was
treated on the basis of thermal vacuum for r-phonons, the above
effects were missed.

The inclusion of the terms $Q^\dag\widetilde{Q}^\dag\ {\rm and}\
\widetilde{Q}^\dag\widetilde{Q}^\dag$ in the trial wave function produces the new poles in the
secular equation (\ref{det}) in comparison with the previous study,\cite{Kos94} namely,
$(\omega_{\lambda_1i_1}-\omega_{\lambda_2i_2})$ and
$-(\omega_{\lambda_1i_1}+\omega_{\lambda_2i_2})$.

The new equations of the TFD-QPM approach (\ref{system2}) and (\ref{det}) are in qualitative
agreement with those in\cite{Seva97} and, in some respects, in.\cite{Bort86}

As it was mentioned in the Introduction, the consideration
in\cite{Bort86,Seva97} was based on the Matsubara Green function
technique and Nuclear Field Theory. The approach of\cite{Seva97}
is especially close to our approach since it treats a hot nucleus
as a system of interacting TRPA phonons. Although the equations of
both the approaches seem to be hardly compared "term-to-term"
because of quite different formalisms explored, in both the cases
the negative TRPA roots are in the game, the poles in the
equations are of the same types, phonon interaction matrix
elements are similar etc. That is why we can establish their
"qualitative" agreement.

The formulae for the matrix elements of $E\lambda$ transitions
from the thermal vacuum state to the state (\ref{trial}) and its
tilde-counterpart are quite obvious. In the leading approximation
the operator $\mathcal{M}(E\lambda)$ induces a transition from the
thermal vacuum to one-phonon components of the thermal state
$|\Psi_\nu(JM)\rangle$ (or $|\widetilde{\Psi}_\nu(JM)\rangle$)
only. Matrix elements of direct transitions from $|0(T);{\rm
RPA}\rangle$ to two-phonon components are very weak although do
not vanish. Thus, one gets
\begin{eqnarray}\label{ampl2}
\Phi(J\nu)&=&\langle\Psi_\nu(JM)\|{\cal M}(E\lambda)\|0(T);{\rm RPA}\rangle=
   \sum_i R_i(J\nu)\Phi_{Ji},
\nonumber\\
\widetilde\Phi(J\nu)&=&\langle\widetilde{\Psi_\nu}(JM)\|{\cal M}(E\lambda)\|0(T);{\rm
RPA}\rangle=\sum_i R_i(J\nu)\widetilde\Phi_{Ji}.
\end{eqnarray}
where $\Phi_{Ji}$ and $\widetilde\Phi_{Ji}$ are given by (\ref{ampl1}).

\section{Summary and conclusions}\label{concl}

The present study was motivated by the necessity to reexamine the
TFD-QPM approach\cite{Kos94,Kos95,Storo04} in theory of hot
nuclei. In this respect, a couple of new facets of the TRPA
formulated within the TFD were found and their influence on the
coupling of thermal phonons was established.

We showed that for the Hamiltonian consisting of a mean field, the
BCS pairing interaction and separable particle-hole effective
interactions, amplitudes of a thermal phonon wave function
(\ref{phonon}) were not determined unambiguously by
diagonalization of the RPA-part of the thermal Hamiltonian. To fix
the coefficients of a linear transformation from the set of
thermal two-quasiparticle operators to phonon operators, one
should impose an additional demand  --- to minimize the
thermodynamic potential of the system of free thermal phonons.
This was achieved by using one more unitary transformation --- a
thermal rotation of "reference" phonons (r-phonons). As a
consequence, the Bose-Einstein thermal occupation numbers (thermal
occupation numbers of phonons) come to play. It should be stressed
that the new ingredient does not affect the main TRPA equation,
i.e., the equation for thermal phonon energies.

As far as we know, the thermal rotation of phonons in the TRPA framework was discussed only
in\cite{Civi93} when treating the pairing BCS Hamiltonian. The thermal rotation of phonons was
missed not only in the preceding TFD-QPM studies\cite{Kos94,Kos95,Storo04} but it was not also
mentioned in.\cite{Tana88,Hats89} Moreover, the Bose-Einstein thermal occupation numbers do not
appear when the TRPA equations are derived by exploring the phonon operator with scattering
terms.\cite{Alas90,Duss90,Civi01}

Thus, according to the present results amplitudes in the thermal
phonon wave function depend not only on thermal occupation numbers
of Bogoliubov quasiparticles making up the phonon but also on the
thermal occupation numbers of the phonon. Moreover, presently the
corresponding thermal phonon vacuum contains some amount of
r-phonons with probabilities determined by their energies in
accordance with the Bose statistics. Due to this, the processes of
excitation and deexcitation of a hot nucleus can be regarded on
equal footing.

The above results have weighty consequences. The first one is the
appearance of the factor $\displaystyle 1/(1- \exp(-\omega/T))$ in
the $E\lambda$ transition strength. Its role was discussed
in\cite{Chom90} where it was derived using the Green function
technique (see also\cite{Som83}). This factor somewhat enhances
the low energy part of the $E\lambda$ transition strength. It is
also essential when $\omega$ can take negative values, i.e.  when
considering a decay of a hot nucleus.

The second consequence concerns the thermal quasiparticle-phonon
interaction $\mathcal{H}_{\rm qph}$. This consequence is two-fold:
a) renormalization of phonon-phonon interaction vertices (cf.
matrix element $U^{\lambda_1i_1}_{\lambda_2i_2}(Ji)$ (\ref{U12})
with that in\cite{Kos94}); b) essential extension of the thermal
two-phonon configuration space and  the corresponding complication
of trial wave function (\ref{trial}).

With the above new ingredients the TFD-QPM approach well conform
(at least qualitatively) with a more traditional one exploring the
Matsubara Green function method.\cite{Bort86,Seva97} Now in both
the approaches one has two kinds of thermal occupation numbers
(fermionic and bosonic) and the identical sets of processes giving
rise to the phonon-phonon coupling.

In view of the above discussion, we conclude that the results of
calculations in\cite{Storo04} should be revised. At the same time,
it should be stressed that the thermal rotation of phonons cannot
be regarded as a mandatory ingredient of a microscopic treatment
of boson-like excitations in many fermion systems. Its necessity
seems to be intimately related with a two-body interaction
character. For example, in
Refs.\cite{Avde96,Storo99,Storo01,Aouissat01,Dzhioev08}, some
approximations going beyond the TRPA were constructed by using the
TFD formalism and their validity was examined by the example of
the Lipkin model. In those studies the transformation from thermal
bifermion operators to thermal phonon operators was unambiguously
determined by diagonalization of the thermal model Hamiltonian. No
additional demands or assumptions were needed to be involved.

One more new feature of the present study is exploring the double
tilde-conjugation rule in the form proposed by
I.~Ojima.\cite{Ojima81} More precisely, we found that just the
Ojima's formulation of DTCR guarantees the correct behaviour of
the thermal phonon wave function in the limit  of vanishing
particle-hole interaction. Furthermore, the Ojima form of DTCR
implies the complex thermal rotation for fermions. An interesting
point is that the effect of the DTCR choice was revealed only at
the TRPA stage while constructing a thermal phonon operator,
whereas at the pure fermionic stage when, e.g., pairing
correlations were considered, the role of DTCR did not show up in
and both the versions gave the same results.

\section{Acknowledgments}

The valuable discussions with Dr. A. Storozhenko and Prof. D.
Kosov are gratefully acknowledged.

\appendix

\section{}

Here we display the matrix elements of the quasiparticle-phonon interaction ${\cal H}_{\rm qph}$
between thermal one-phonon  and different two-phonon configurations. The phonon amplitudes are
shown in \eqref{amplitudes} and include both quasiparticle and phonon occupation factors. The
symbol $\Bigl\{\cdots\Bigr\}$ is the standard $6j$-symbol.
 \begin{multline}\label{U12}
 U^{\lambda_1i_1}_{\lambda_2i_2}(Ji)=
 \hat{\lambda}_1\hat{\lambda}_2\sum_\tau{\sum_{j_1j_2j_3}}^{\tau}\left[
\frac{f^{(J)}_{j_1j_2}}{\sqrt{{\cal N}^{Ji}_\tau}}
 \left\{\begin{array}{ccc}J&\lambda_1&\lambda_2\\j_3&j_1&j_2\end{array}\right\}
 {\cal G}^{Ji}_{\lambda_1i_1\lambda_2i_2}(j_1j_2j_3)\right. \\
 +\left. \frac{f^{(\lambda_1)}_{j_1j_2}}{\sqrt{{\cal N}^{\lambda_1i_1}_\tau}}
 \left\{\begin{array}{ccc}\lambda_1&\lambda_2&J\\j_3&j_1&j_2\end{array}\right\}
 {\cal F}^{\lambda_1i_1}_{\lambda_2i_2Ji}(j_1j_2j_3)\right. \\
\left. +(-1)^{\lambda_1+\lambda_2+J}
 \frac{f^{(\lambda_2)}_{j_1j_2}}{\sqrt{{\cal N}^{\lambda_2i_2}_\tau}}
 \left\{\begin{array}{ccc}\lambda_2&\lambda_1&J\\j_3&j_1&j_2\end{array}\right\}
  {\cal F}^{\lambda_2i_2}_{\lambda_1i_1 Ji}(j_1j_2j_3)\right],
 \end{multline}
where
\begin{eqnarray*}
    {\cal G}^{Ji}_{\lambda_1i_1\lambda_2i_2}(j_1j_2j_3)&=&
  {\cal X}^{Ji}_{j_1j_2} v^{(-)}_{j_1j_2}
  \bigl(\psi^{\lambda_1i_1}_{j_2j_3}\phi^{\lambda_2i_2}_{j_3j_1}+
   \phi^{\lambda_1i_1}_{j_2j_3}\psi^{\lambda_2i_2}_{j_3j_1}+
   \eta^{\lambda_1i_1}_{j_2j_3}\widetilde\xi^{\lambda_2i_2}_{j_3j_1}+
   \xi^{\lambda_1i_1}_{j_2j_3}\widetilde\eta^{\lambda_2i_2}_{j_3j_1}\bigr)\\
   &-&{\cal Y}^{Ji}_{j_1j_2} v^{(-)}_{j_1j_2}
  \bigl(\widetilde\psi^{\lambda_1i_1}_{j_2j_3}\widetilde\phi^{\lambda_2i_2}_{j_3j_1}+
   \widetilde\phi^{\lambda_1i_1}_{j_2j_3}\widetilde\psi^{\lambda_2i_2}_{j_3j_1}+
   \widetilde\eta^{\lambda_1i_1}_{j_2j_3}\xi^{\lambda_2i_2}_{j_3j_1}+
   \widetilde\xi^{\lambda_1i_1}_{j_2j_3}\eta^{\lambda_2i_2}_{j_3j_1}\bigl)\\
  &+&{\cal Z}^{Ji}_{j_1j_2} u^{(+)}_{j_1j_2}
    \bigl(\widetilde{\psi}^{\lambda_1i_1}_{j_2j_3}\widetilde\xi^{\lambda_2i_2}_{j_3j_1}+
   \widetilde{\phi}^{\lambda_1i_1}_{j_2j_3}\widetilde\eta^{\lambda_2i_2}_{j_3j_1}-
   \widetilde\eta^{\lambda_1i_1}_{j_2j_3}\phi^{\lambda_2i_2}_{j_3j_1}-
   \widetilde\xi^{\lambda_1i_1}_{j_2j_3}\psi^{\lambda_2i_2}_{j_3j_1}\bigr)\\
  &-&{\cal Z}^{Ji}_{j_2j_1} u^{(+)}_{j_1j_2}
   \bigl(\psi^{\lambda_1i_1}_{j_2j_3}\xi^{\lambda_2i_2}_{j_3j_1}+
    \phi^{\lambda_1i_1}_{j_2j_3}\eta^{\lambda_2i_2}_{j_3j_1}-
    \eta^{\lambda_1i_1}_{j_2j_3}\widetilde\phi^{\lambda_2i_2}_{j_3j_1}-
   \xi^{\lambda_1i_1}_{j_2j_3}\widetilde\psi^{\lambda_2i_2}_{j_3j_1}\bigr)~,\\[5mm]
    {\cal F}^{\lambda_1i_1}_{\lambda_2i_2Ji}(j_1j_2j_3)&=&
   {\cal X}^{\lambda_1i_1}_{j_1j_2} v^{(-)}_{j_1j_2}
  \bigl(\psi^{\lambda_2i_2}_{j_2j_3}\psi^{Ji}_{j_3j_1}+
   \phi^{\lambda_2i_2}_{j_2j_3}\phi^{Ji}_{j_3j_1}+
   \eta^{\lambda_2i_2}_{j_2j_3}\widetilde\eta^{Ji}_{j_3j_1}+
    \xi^{\lambda_2i_2}_{j_2j_3}\widetilde\xi^{Ji}_{j_3j_1}\bigr)\\
  &-&{\cal Y}^{\lambda_1i_1}_{j_1j_2} v^{(-)}_{j_1j_2}
  \bigl(\widetilde\psi^{\lambda_2i_2}_{j_2j_3}\widetilde\psi^{Ji}_{j_3j_1}+
   \widetilde\phi^{\lambda_2i_2}_{j_2j_3}\widetilde\phi^{Ji}_{j_3j_1}+
   \widetilde\eta^{\lambda_2i_2}_{j_2j_3}\eta^{Ji}_{j_3j_1}+
    \widetilde\xi^{\lambda_2i_2}_{j_2j_3}\xi^{Ji}_{j_3j_1}\bigr)\\
 &+&{\cal Z}^{\lambda_1i_1}_{j_1j_2} u^{(+)}_{j_1j_2}
  \bigl(\widetilde{\psi}^{\lambda_2i_2}_{j_2j_3}\widetilde\eta^{Ji}_{j_3j_1}+
   \widetilde{\phi}^{\lambda_2i_2}_{j_2j_3}\widetilde\xi^{Ji}_{j_3j_1}-
   \widetilde\eta^{\lambda_2i_2}_{j_2j_3}\psi^{Ji}_{j_3j_1}-
   \widetilde\xi^{\lambda_2i_2}_{j_2j_3}\phi^{Ji}_{j_3j_1}\bigr)\\
  &-&{\cal Z}^{\lambda_1i_1}_{j_2j_1}u^{(+)}_{j_1j_2}
   \bigl(\psi^{\lambda_2i_2}_{j_2j_3}\eta^{Ji}_{j_3j_1}+
    \phi^{\lambda_2i_2}_{j_2j_3}\xi^{Ji}_{j_3j_1}-
    \eta^{\lambda_2i_2}_{j_2j_3}\widetilde\psi^{Ji}_{j_3j_1}-
    \xi^{\lambda_2i_2}_{j_2j_3}\widetilde\phi^{Ji}_{j_3j_1}\bigr)~.
     \end{eqnarray*}

 \begin{multline}\label{V12}
   V^{\lambda_1i_1}_{\lambda_2i_2}(Ji)=
 \hat{\lambda}_1\hat{\lambda}_2\sum_\tau{\sum_{j_1j_2j_3}}^{\tau}\left[
\frac{f^{(J)}_{j_1j_2}}{\sqrt{{\cal N}^{Ji}_\tau}}
 \left\{\begin{array}{ccc}J&\lambda_1&\lambda_2\\j_3&j_1&j_2\end{array}\right\}
 {\cal S}^{Ji}_{\lambda_1i_1\lambda_2i_2}(j_1j_2j_3)\right. \\
\left. +\frac{f^{(\lambda_1)}_{j_1j_2}}{\sqrt{{\cal N}^{\lambda_1i_1}_\tau}}
 \left\{\begin{array}{ccc}\lambda_1&\lambda_2&J\\j_3&j_1&j_2\end{array}\right\}
 \widetilde{\cal R}^{\lambda_1i_1}_{\lambda_2i_2Ji}(j_1j_2j_3) \right. \\
\left. -(-1)^{\lambda_1+\lambda_2+J}\frac{f^{(\lambda_2)}_{j_1j_2}}{\sqrt{{\cal
N}^{\lambda_2i_2}_\tau}}
 \left\{\begin{array}{ccc}\lambda_2&\lambda_1&J\\j_3&j_1&j_2\end{array}\right\}
  \widetilde{\cal F}^{\lambda_2i_2}_{\lambda_1i_1 Ji}(j_1j_2j_3)\right],
   \end{multline}
  where
  \begin{eqnarray*}
    {\cal S}^{Ji}_{\lambda_1i_1\lambda_2i_2}(j_1j_2j_3)&=&
    {\cal X}^{Ji}_{j_1j_2}v^{(-)}_{j_1j_2}
   \bigl(\psi^{\lambda_1i_1}_{j_2j_3}\widetilde\phi^{\lambda_2i_2}_{j_3j_1}+
   \phi^{\lambda_1i_1}_{j_2j_3}\widetilde\psi^{\lambda_2i_2}_{j_3j_1}+
   \eta^{\lambda_1i_1}_{j_2j_3}\xi^{\lambda_2i_2}_{j_3j_1}+
   \xi^{\lambda_1i_1}_{j_2j_3}\eta^{\lambda_2i_2}_{j_3j_1}\bigr)\\
   &-&{\cal Y}^{Ji}_{j_1j_2}v^{(-)}_{j_1j_2}
  \bigl(\widetilde\psi^{\lambda_1i_1}_{j_2j_3}\phi^{\lambda_2i_2}_{j_3j_1}+
   \widetilde\phi^{\lambda_1i_1}_{j_2j_3}\psi^{\lambda_2i_2}_{j_3j_1}+
   \widetilde\eta^{\lambda_1i_1}_{j_2j_3}\widetilde\xi^{\lambda_2i_2}_{j_3j_1}+
   \widetilde\xi^{\lambda_1i_1}_{j_2j_3}\widetilde\eta^{\lambda_2i_2}_{j_3j_1}\bigr)\\
   &+&{\cal Z}^{Ji}_{j_1j_2}u^{(+)}_{j_1j_2}
  \bigl(\widetilde{\psi}^{\lambda_1i_1}_{j_2j_3}\xi^{\lambda_2i_2}_{j_3j_1}+
   \widetilde{\phi}^{\lambda_1i_1}_{j_2j_3}\eta^{\lambda_2i_2}_{j_3j_1}-
   \widetilde\eta^{\lambda_1i_1}_{j_2j_3}\widetilde\phi^{\lambda_2i_2}_{j_3j_1}-
   \widetilde\xi^{\lambda_1i_1}_{j_2j_3}\widetilde\psi^{\lambda_2i_2}_{j_3j_1}\bigr)\\
   &-&{\cal Z}^{Ji}_{j_2j_1}u^{(+)}_{j_1j_2}
   \bigl(\psi^{\lambda_1i_1}_{j_2j_3}\widetilde\xi^{\lambda_2i_2}_{j_3j_1}+
    \phi^{\lambda_1i_1}_{j_2j_3}\widetilde\eta^{\lambda_2i_2}_{j_3j_1}-
    \eta^{\lambda_1i_1}_{j_2j_3}\phi^{\lambda_2i_2}_{j_3j_1}-
   \xi^{\lambda_1i_1}_{j_2j_3}\psi^{\lambda_2i_2}_{j_3j_1}\bigr)~,\\[5mm]
    \widetilde{\cal R}^{\lambda_1i_1}_{\lambda_2i_2Ji}(j_1j_2j_3)&=&
  {\cal X}^{\lambda_1i_1}_{j_1j_2}v^{(-)}_{j_1j_2}
  \bigl(\widetilde\psi^{\lambda_2i_2}_{j_2j_3}\psi^{Ji}_{j_3j_1}+
   \widetilde\phi^{\lambda_2i_2}_{j_2j_3}\phi^{Ji}_{j_3j_1}+
   \widetilde\eta^{\lambda_2i_2}_{j_2j_3}\widetilde\eta^{Ji}_{j_3j_1}+
    \widetilde\xi^{\lambda_2i_2}_{j_2j_3}\widetilde\xi^{Ji}_{j_3j_1}\bigr)\\
   &-&{\cal Y}^{\lambda_1i_1}_{j_1j_2}v^{(-)}_{j_1j_2}
  \bigl(\psi^{\lambda_2i_2}_{j_2j_3}\widetilde\psi^{Ji}_{j_3j_1}+
   \phi^{\lambda_2i_2}_{j_2j_3}\widetilde\phi^{Ji}_{j_3j_1}+
   \eta^{\lambda_2i_2}_{j_2j_3}\eta^{Ji}_{j_3j_1}+
   \xi^{\lambda_2i_2}_{j_2j_3}\xi^{Ji}_{j_3j_1}\bigr)\\
   &+&{\cal Z}^{\lambda_1i_1}_{j_1j_2}u^{(+)}_{j_1j_2}
  \bigl(\psi^{\lambda_2i_2}_{j_2j_3}\widetilde\eta^{Ji}_{j_3j_1}+
   \phi^{\lambda_2i_2}_{j_2j_3}\widetilde\xi^{Ji}_{j_3j_1}-
   \eta^{\lambda_2i_2}_{j_2j_3}\psi^{Ji}_{j_3j_1}-
   \xi^{\lambda_2i_2}_{j_2j_3}\phi^{Ji}_{j_3j_1}\bigr)\\
   &-&{\cal Z}^{\lambda_1i_1}_{j_2j_1}u^{(+)}_{j_1j_2}
   \bigl(\widetilde\psi^{\lambda_2i_2}_{j_2j_3}\eta^{Ji}_{j_3j_1}+
    \widetilde\phi^{\lambda_2i_2}_{j_2j_3}\xi^{Ji}_{j_3j_1}-
    \widetilde\eta^{\lambda_2i_2}_{j_2j_3}\widetilde\psi^{Ji}_{j_3j_1}-
    \widetilde\xi^{\lambda_2i_2}_{j_2j_3}\widetilde\phi^{Ji}_{j_3j_1}\bigr)~,\\[5mm]
  \widetilde{\cal F}^{\lambda_2i_2}_{\lambda_1i_1 Ji}(j_1j_2j_3)&=&
 {\cal X}^{\lambda_2i_2}_{j_1j_2}v^{(-)}_{j_1j_2}
  \bigl(\widetilde\psi^{\lambda_1i_1}_{j_2j_3}\widetilde\psi^{Ji}_{j_3j_1}+
   \widetilde\phi^{\lambda_1i_1}_{j_2j_3}\widetilde\phi^{Ji}_{j_3j_1}+
   \widetilde\eta^{\lambda_1i_1}_{j_2j_3}\eta^{Ji}_{j_3j_1}+
   \widetilde \xi^{\lambda_1i_1}_{j_2j_3}\xi^{Ji}_{j_3j_1}\bigr)\\
  &-&{\cal Y}^{\lambda_2i_2}_{j_1j_2}v^{(-)}_{j_1j_2}
  \bigl(\psi^{\lambda_1i_1}_{j_2j_3}\psi^{Ji}_{j_3j_1}+
   \phi^{\lambda_1i_1}_{j_2j_3}\phi^{Ji}_{j_3j_1}+
   \eta^{\lambda_2i_3}_{j_3j_1}\widetilde\eta^{Ji}_{j_3j_1}+
    \xi^{\lambda_2i_3}_{j_3j_1}\widetilde\xi^{Ji}_{j_3j_1}\bigr)\\
  &+&{\cal Z}^{\lambda_2i_2}_{j_1j_2}u^{(+)}_{j_1j_2}
  \bigl(\psi^{\lambda_1i_1}_{j_2j_3}\eta^{Ji}_{j_3j_1}+
   \phi^{\lambda_1i_1}_{j_2j_3}\xi^{Ji}_{j_3j_1}-
   \eta^{\lambda_1i_1}_{j_2j_3}\widetilde\psi^{Ji}_{j_3j_1}-
    \xi^{\lambda_1i_1}_{j_2j_3}\widetilde\phi^{Ji}_{j_3j_1}\bigr)\\
   &-&{\cal Z}^{\lambda_2i_2}_{j_2j_1}u^{(+)}_{j_1j_2}
   \bigl(\widetilde\psi^{\lambda_1i_1}_{j_2j_3}\widetilde\eta^{Ji}_{j_3j_1}+
   \widetilde\phi^{\lambda_1i_1}_{j_2j_3}\widetilde\xi^{Ji}_{j_3j_1}-
   \widetilde\eta^{\lambda_1i_1}_{j_2j_3}\psi^{Ji}_{j_3j_1}-
   \widetilde \xi^{\lambda_1i_1}_{j_2j_3}\phi^{Ji}_{j_3j_1}\bigr)~.
   \end{eqnarray*}

 \begin{multline}\label{W12}
 W^{\lambda_1i_1}_{\lambda_2i_2}(Ji)=
\hat{\lambda}_1\hat{\lambda}_2\sum_\tau{\sum_{j_1j_2j_3}}^{\tau}\left[
\frac{f^{(J)}_{j_1j_2}}{\sqrt{{\cal N}^{Ji}_\tau}}
 \left\{\begin{array}{ccc}J&\lambda_1&\lambda_2\\j_3&j_1&j_2\end{array}\right\}
 \widetilde{\cal G}^{Ji}_{\lambda_1i_1\lambda_2i_2}(j_1j_2j_3)\right. \\
\left. -\frac{f^{(\lambda_1)}_{j_1j_2}}{\sqrt{{\cal N}^{\lambda_1i_1}_\tau}}
 \left\{\begin{array}{ccc}\lambda_1&\lambda_2&J\\j_3&j_1&j_2\end{array}\right\}
 {\cal R}^{\lambda_1i_1}_{\lambda_2i_2Ji}(j_1j_2j_3) \right. \\
 \left. -(-1)^{\lambda_1+\lambda_2+J}
 \frac{f^{(\lambda_2)}_{j_1j_2}}{\sqrt{{\cal N}^{\lambda_2i_2}_\tau}}
 \left\{\begin{array}{ccc}\lambda_2&\lambda_1&J\\j_3&j_1&j_2\end{array}\right\}
  {\cal R}^{\lambda_2i_2}_{\lambda_1i_1Ji}(j_1j_2j_3)\right],
 \end{multline}
 where

\begin{eqnarray*}
   \widetilde{\cal G}^{Ji}_{\lambda_1i_1\lambda_2i_2}(j_1j_2j_3)&=&
   {\cal X}^{Ji}_{j_1j_2} v^{(-)}_{j_1j_2}
  (\widetilde\psi^{\lambda_1i_1}_{j_2j_3}\widetilde\phi^{\lambda_2i_2}_{j_3j_1}+
   \widetilde\phi^{\lambda_1i_1}_{j_2j_3}\widetilde\psi^{\lambda_2i_2}_{j_3j_1}+
   \widetilde\eta^{\lambda_1i_1}_{j_2j_3}\xi^{\lambda_2i_2}_{j_3j_1}+
   \widetilde\xi^{\lambda_1i_1}_{j_2j_3}\eta^{\lambda_2i_2}_{j_3j_1})\\
  &-&{\cal Y}^{Ji}_{j_1j_2} v^{(-)}_{j_1j_2}
  (\psi^{\lambda_1i_1}_{j_2j_3}\phi^{\lambda_2i_2}_{j_3j_1}+
   \phi^{\lambda_1i_1}_{j_2j_3}\psi^{\lambda_2i_2}_{j_3j_1}+
   \eta^{\lambda_1i_1}_{j_2j_3}\widetilde\xi^{\lambda_2i_2}_{j_3j_1}+
   \xi^{\lambda_1i_1}_{j_2j_3}\widetilde\eta^{\lambda_2i_2}_{j_3j_1})\\
   &+&{\cal Z}^{Ji}_{j_1j_2} u^{(+)}_{j_1j_2}
   (\psi^{\lambda_1i_1}_{j_2j_3}\xi^{\lambda_2i_2}_{j_3j_1}+
    \phi^{\lambda_1i_1}_{j_2j_3}\eta^{\lambda_2i_2}_{j_3j_1}-
    \eta^{\lambda_1i_1}_{j_2j_3}\widetilde\phi^{\lambda_2i_2}_{j_3j_1}-
   \xi^{\lambda_1i_1}_{j_2j_3}\widetilde\psi^{\lambda_2i_2}_{j_3j_1})\\
   &-&{\cal Z}^{Ji}_{j_1j_2} u^{(+)}_{j_1j_2}
   (\widetilde{\psi}^{\lambda_1i_1}_{j_2j_3}\widetilde\xi^{\lambda_2i_2}_{j_3j_1}+
   \widetilde{\phi}^{\lambda_1i_1}_{j_2j_3}\widetilde\eta^{\lambda_2i_2}_{j_3j_1}-
   \widetilde\eta^{\lambda_1i_1}_{j_2j_3}\phi^{\lambda_2i_2}_{j_3j_1}-
   \widetilde\xi^{\lambda_1i_1}_{j_2j_3}\psi^{\lambda_2i_2}_{j_3j_1})~,\\[5mm]
    {\cal R}^{\lambda_1i_1}_{\lambda_2i_2Ji}(j_1j_2j_3)&=&
 {\cal X}^{\lambda_1i_1}_{j_1j_2}v^{(-)}_{j_1j_2}
  (\psi^{\lambda_2i_2}_{j_2j_3}\widetilde\psi^{Ji}_{j_3j_1}+
   \phi^{\lambda_2i_2}_{j_2j_3}\widetilde\phi^{Ji}_{j_3j_1}+
   \eta^{\lambda_2i_2}_{j_2j_3}\eta^{Ji}_{j_3j_1}+
   \xi^{\lambda_2i_2}_{j_2j_3}\xi^{Ji}_{j_3j_1})\\
   &-&{\cal Y}^{\lambda_1i_1}_{j_1j_2}v^{(-)}_{j_1j_2}
  (\widetilde\psi^{\lambda_2i_2}_{j_2j_3}\psi^{Ji}_{j_3j_1}+
   \widetilde\phi^{\lambda_2i_2}_{j_2j_3}\phi^{Ji}_{j_3j_1}+
   \widetilde\eta^{\lambda_2i_2}_{j_2j_3}\widetilde\eta^{Ji}_{j_3j_1}+
    \widetilde\xi^{\lambda_2i_2}_{j_2j_3}\widetilde\xi^{Ji}_{j_3j_1})\\
   &+&{\cal Z}^{\lambda_1i_1}_{j_1j_2}u^{(+)}_{j_1j_2}
  (\widetilde\psi^{\lambda_2i_2}_{j_2j_3}\eta^{Ji}_{j_3j_1}+
    \widetilde\phi^{\lambda_2i_2}_{j_2j_3}\xi^{Ji}_{j_3j_1}-
    \widetilde\eta^{\lambda_2i_2}_{j_2j_3}\widetilde\psi^{Ji}_{j_3j_1}-
    \widetilde\xi^{\lambda_2i_2}_{j_2j_3}\widetilde\phi^{Ji}_{j_3j_1})\\
  &-&{\cal Z}^{\lambda_1i_1}_{j_2j_1}u^{(+)}_{j_1j_2}
   (\psi^{\lambda_2i_2}_{j_2j_3}\widetilde\eta^{Ji}_{j_3j_1}+
   \phi^{\lambda_2i_2}_{j_2j_3}\widetilde\xi^{Ji}_{j_3j_1}-
   \eta^{\lambda_2i_2}_{j_2j_3}\psi^{Ji}_{j_3j_1}-
   \xi^{\lambda_2i_2}_{j_2j_3}\phi^{Ji}_{j_3j_1})~.
     \end{eqnarray*}

\eject

\end{document}